\documentclass[%
 aip,
 amsmath,amssymb,
 reprint
]{revtex4-1}
\usepackage{natbib}

\usepackage{siunitx}
\usepackage{array}
\usepackage{subfig}
\usepackage{bm}
\usepackage{booktabs}
\usepackage{xcolor}
\usepackage{moreverb}
\setcitestyle{numbers,open={[},close={]}}
\usepackage{amssymb}
\usepackage{amsmath}
\usepackage{mathtools}
\usepackage{multirow}
\usepackage{array}
\usepackage{wasysym}
\usepackage{booktabs}
\usepackage{gensymb}
\usepackage{caption}
\usepackage{graphicx}
\usepackage{tabularx}
\usepackage{stackengine}
\usepackage{pbox}

\usepackage[margin=2.5cm]{geometry}
\usepackage{xcolor}
\usepackage{bm}%
\usepackage[colorlinks=true,linkcolor=blue,citecolor=red]{hyperref}%
\expandafter\ifx\csname package@font\endcsname\relax\else
\expandafter\expandafter
\expandafter\usepackage
\expandafter\expandafter
\expandafter{\csname package@font\endcsname}%
\fi
\graphicspath{{./fig/}}

\begin{document}

\title{Microstructural Smoothed Particle Hydrodynamics Model and Simulations of Discontinuous Shear-Thickening Fluids}%
\author{P. Angerman}%
\email{2037484@swansea.ac.uk}
\affiliation{Complex Fluids Research Group, Department of Chemical Engineering, Swansea University, Swansea SA1 8EN, United Kingdom}
\affiliation{Basque Center for Applied Mathematics (BCAM), Alameda de Mazarredo 14, 48009 Bilbao, Spain}%
\date{}%
\author{S.S. Prasanna Kumar}%
\affiliation{Basque Center for Applied Mathematics (BCAM), Alameda de Mazarredo 14, 48009 Bilbao, Spain}%
\author{R. Seto}%
\affiliation{Wenzhou Key Laboratory of Biomaterials and Engineering, Wenzhou Institute, University of Chinese Academy of Sciences, Wenzhou, 325000, China}
\affiliation{Oujiang Laboratory (Zhejiang Lab for Regenerative Medicine, Vision and Brain Health), Wenzhou, 325000, China}
\affiliation{Graduate School of Information Science, University of Hyogo, 
Kobe, 650-0047, Japan}

\author{B. Sandnes}%
\affiliation{Complex Fluids Research Group, Department of Chemical Engineering, Swansea University, Swansea SA1 8EN, United Kingdom}
\author{M. Ellero}%

\affiliation{Basque Center for Applied Mathematics (BCAM), Alameda de Mazarredo 14, 48009 Bilbao, Spain}
\affiliation{IKERBASQUE, Basque Foundation for Science, Calle de Mar\'{i}a D\'{i}az de Haro 3, 48013 Bilbao, Spain}
\affiliation{Zienkiewicz Centre for Computational Engineering (ZCCE), Swansea University, Bay Campus, Swansea SA1 8EN, United Kingdom}

\revised{}%
\begin{abstract}
Despite the recent interest in the discontinuous shear-thickening (DST) behaviour, few computational
works tackle the rich hydrodynamics of these fluids. In this work, we present the first implementation
of a microstructural DST model in Smoothed Particle Hydrodynamic (SPH) simulation. The scalar model was
implemented in an SPH scheme and tested in two flow geometries. Three distinct ratios of local to non-local microstructural effects were probed: weak, moderate, and strong non-locality. Strong and moderate cases yielded excellent agreement with flow curves constructed via the Wyart--Cates (WC) model, with the moderate case exhibiting banding patterns. Weak non-locality produced stress-splitting instability, resulting in discontinuous stress fields and poor agreement with the WC model. The mechanism of the stress-splitting has been explored and contextualised by the interaction of local microstructure evolution and the stress-control scheme. Velocity profiles
obtained in body force-driven channel flow were found to be in excellent agreement with the
analytical solution, yielding an upward inflection corresponding to the typical S-curve. Simulations
carried out at increasing driving forces exhibited a decrease in flow. We showed that even
the simple scalar model can capture some of the key properties of DST materials, laying the foundation
for further SPH study of instabilities and pattern formation.
\end{abstract}
\maketitle

\section{Introduction}
Suspensions of both colloidal\,\cite{Mewis} and non-colloidal\,\cite{TannerReview,MortonReview} particles are ubiquitous in industrially relevant flows\,\cite{tanner2000engineering}, and their rich properties have been subject of extensive study. 
Some dense suspensions of fine non-Brownian particles can exhibit discontinuous shear thickening (DST) behaviour\,\cite{MorrisReview}, which has been a subject of great interest in the last decade. 
With possible implications in industries spanning from manufacturing, environmental engineering, and military application, understanding the underlying mechanism responsible for the unique behaviour has become paramount. 
Multiple physical explanations have been proposed\,\cite{Brown_2014}, with the most widely adopted being the mechanism of \citet{MariSeto} or Wyart--Cates (WC) model\,\cite{WC2014}, where the particles are held apart in a frictionless state by a characteristic repulsive force. 
Upon application of external stress, the interparticle repulsion is overcome, bringing about a transition to a frictional state where the developed contact network inhibits the ease of motion of the suspension.
A defining feature of DST, as opposed to continuous shear-thickening (CST), is the characteristic `S' shape of the flow curve. 
This unique non-monotonic shape facilitates unstable behaviours in rheometric\,\cite{Herle,Richards} and complex flows.
For example, Texier et al.\,\cite{Texier2020,Texier2023}
demonstrated that the negative gradients of flow curve ($\frac{d\dot{\gamma}}{d\sigma}<0$) present in DST materials cause instability in free surface flows of cornstarch suspension, leading to roll-waves.

WC-type models are inherently microstructural --- the key parameter governing the state of the suspension at any point is the arrangement of the particle contact network. 
Typically, works in the context of WC theory propose a relationship between development of the microstructure and macro-hydrodynamic flow properties through a transient evolution equation.
\citet{Nakanishi} employed a scalar model with an assumed S-shape rheology curve
in a range of geometries, including simple shear, Poiseuille flow, and free surface inclined plane. 
The results in simple shear exhibited periodic solutions, accompanied by formation of inhomogeneous fields of the microstructural parameter across the gap.
\citet{Kamrin2019} proposed a scalar model based on the balance of strain-dependant hardening and softening attributed to load buckling, shearing, and electrostatic repulsion. 
Their model predicted rheological behaviour in good agreement with the experimental evidence. 

Another development comes from the work of \citet{WilsonCates2019}, who proposed a full tensorial evolution equation. 
In addition to predicting DST rheology, they were able to predict negligible first normal stress difference, significant negative second normal stress difference, and flow behaviour in more complicated flow arrangements, including flow reversal and superposed transverse oscillations\,\cite{WilsonCates2020}, although the values of normal stress differences predicted by their model were quantitatively too high ($N_1$) and too low ($N_2$) relative to the DEM values, along with the increases during thickening being underestimated.
Their model also did not consider the role of contact friction in the development of microstructure and excluded tangential forces and torque balance.

%


So far, nearly all continuum microstructural models have been implemented in Eulerian frameworks. 
In recent times, there has been much interest in simulations of complex fluids with grid-less methods, primarily Smoothed Particle Hydrodynamics (SPH)\,\cite{Lind2020, Ellero2007}. 
SPH was originally developed by \citet{Monaghan1977} and \citet{Lucy1977} in 1977 in the study of astrophysical problems. 
It is a fully Lagrangian scheme, where the fluid is represented by a set of discrete `particles' carried by flow. 
Hydrodynamic properties are computed at the position of each `particle' by averaging via a smoothing kernel at each time step, and the trajectories of particles evolve over time.
In comparison with Eulerian and mixed Lagrangian--Eulerian methods, SPH has a number of advantages and disadvantages.
Generally, grid-based schemes benefit from superior convergence rates and easier implementation of boundary conditions\,\cite{Lind2020}.
On the other hand, SPH is inherently conservative and handles free surface boundaries and interfacial multiphase flows naturally\,\cite{Monaghan2005} ---
which is of importance in a number of studies, including the behaviour of granular bed loads\,\cite{Shimizu2016}, mixing\,\cite{Reece2020}, and cavitating flows\,\cite{KALATEH202051}. 
Moreover, SPH allows for direct access to the flow history --- a crucial feature for microstructural models of complex fluids. 
This can be leveraged in heterogeneous multi-scale modeling (HMM)\,\cite{MorenoGeneral}, with an example application in biologically relevant flows\,\cite{moreno2023morphological}, or by directly solving integral constitutive equations\,\cite{Luca2023UR}.

Beyond Newtonian fluids, SPH has been used in a host of fields, including solid and complex fluid mechanics. 
In 2002, \citet{Ellero2002} simulated viscoelastic flow by incorporating the corotational Maxwell model in SPH, followed by Oldroyd-B\,\cite{Tanner2005}.
Since then, Fang et al.\,\cite{Fang2006, Fang2009} studied the behaviour of free surface flows of an Oldroyd-B fluid, with a focus on addressing tensile instability.
\citet{Hosseini2007} implemented both Maxwell and Oldroyd-B models in SPH free surface flows, including impacting drop and jet buckling. 
\citet{Liu2012} simulated 3D injection molding of a Cross-model fluid. 
Implementation of Oldroyd-B was studied in a periodic array of cylinders by \citet{Ellero2012} and \citet{Muzio}, and in extrudate swell by \citet{Xu2016}. 

Furthermore, 
a range of viscoplastic and elasto-viscoplastic models have been implemented in SPH. 
\citet{BonetRodriguez-Paz} simulated debris flow down an incline using the Bingham model and the generalized viscoplastic model. 
\citet{MinattiParis} considered free surface granular flows, with a constitutive relation based on the works of \citet{Pouliquen_2006} and \citet{Jop2006}, 
and validated their model against granular column collapse experiments. 
Furthermore, Bingham-like viscoplastic models have been applied to sedimenting flows\,\cite{Fourtaka,Khanpour}, rheometric flows\,\cite{ZHU2010362} and mixed fluid-structure cases\,\cite{PAIVA2009306,Zhu2018}. 
Recently, \citet{Rossi2022} implemented a modified microstructural Papanastasiou model to study thixo-viscoplastic flows around a cylinder. 
The first application of SPH to DST flow was carried out in the work of \citet{Wagner2017}, wherein they considered a non-microstructural inverse biviscous model in a simple planar flow.


In this work, we present in detail the first implementation of a scalar microstructural DST rheological model in an SPH scheme. 
Unlike previous models\,\cite{Wagner2017}, we obtain typical S-shaped DST rheology in the simple shear and characteristically inflexed velocity profiles accompanied by flow reduction behaviour in channel flow. 
In both cases, stress-splitting is observed. 
Section\,\ref{sec_Governing_Equations} lays out the details of the model used in this study.
Section\,\ref{sec_SPH-DST_model}
presents the details of the numerical scheme and boundary condition treatment. 
Finally, results for simple shear and planar channel flow are presented 
in sections\,\ref{subsec_Stress-Imposed_Shear_Flow}
 and \ref{subsec_Channel_flow}, respectively.

\section{Governing Equations}
\label{sec_Governing_Equations}

In this section, we outline the main equations used to model our DST fluid system.
The continuum system can be written in the Lagrangian frame:
\begin{gather}
 \frac{d\rho}{dt}=-\rho\nabla \cdot \boldsymbol{v}, \\
\frac{d\boldsymbol{v}}{dt}=\frac{1}{\rho}\nabla\cdot\boldsymbol{\sigma}+\boldsymbol{F}, 
\label{momentum_equation}
\\ 
 \boldsymbol{\sigma}=-p\boldsymbol{I}+
 \eta \left(\nabla \boldsymbol{v}+\nabla\boldsymbol{v^T}\right),
\end{gather}
where $d/dt \equiv \partial/\partial t + \boldsymbol{v}\cdot\nabla$ is the material derivative,  $\rho(\boldsymbol{x},t)$ is the density field, $\boldsymbol{v}(\boldsymbol{x},t)$ is the velocity field, $\boldsymbol{\sigma}(\boldsymbol{x},t)$ is the stress field, $p$ is the pressure, and $\boldsymbol{F}$ is a body force acting on fluid. In this DST model, the local suspension viscosity is determined using the Maron--Pierce 
expression\,\cite{Maron}.
\begin{equation}
\label{Maron–Pierce}
  \eta(\phi,\phi^{\mathrm{J}},t)=\eta_{\mathrm{s}}
  \left(1-\frac{\phi}{\phi^{\mathrm{J}} } \right)^{-2},
\end{equation}
where $\eta(\phi,\phi^{\mathrm{J}},t)$ and $\eta_{\mathrm{s}} $ are, respectively, the suspension and constant solvent viscosities. 
$\phi$ is the particle volume fraction, and $\phi^{\mathrm{J}}$ is the jamming volume fraction. 
Assuming particle migration is negligible for the problems studied in the present work, the value of $\phi$ is kept constant. 
The value of $\phi^{\mathrm{J}}$ is inferred dynamically as
\begin{equation}
\label{jamming_point}
 \phi^{\mathrm{J}} (f(\boldsymbol{x},t))=\phi^{\mathrm{m}} 
 f(\boldsymbol{x},t)+\phi^0
 \{ 1-f(\boldsymbol{x},t) \},
\end{equation}
where $\phi^{\mathrm{m}}$ and $\phi^0$ are frictional and frictionless divergence volume fractions, respectively.

\begin{figure*}[tbh]
\centering
\subfloat[][]{\includegraphics[width=0.48\textwidth]{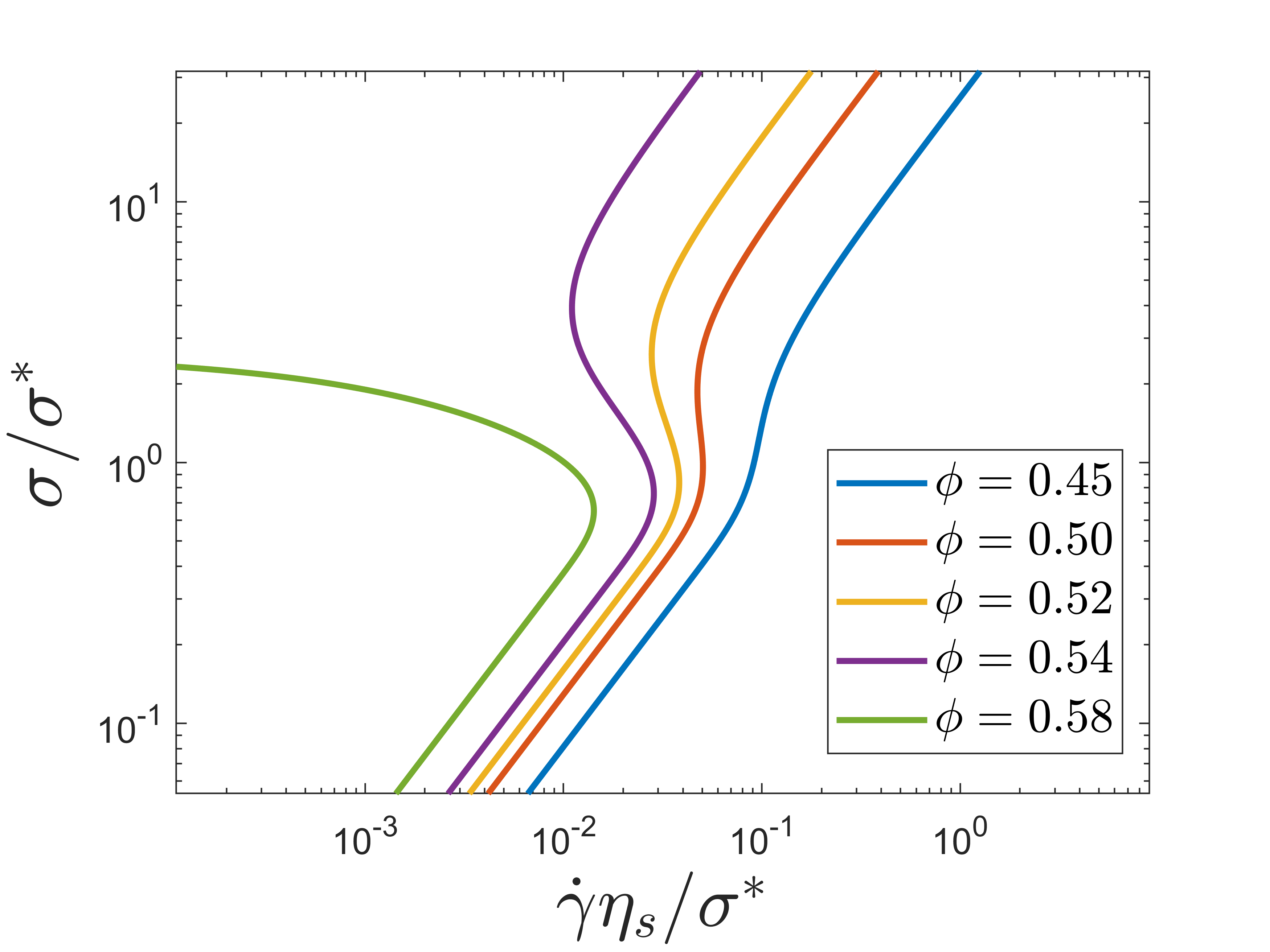}}
\subfloat[][]{\includegraphics[width=0.48\textwidth]{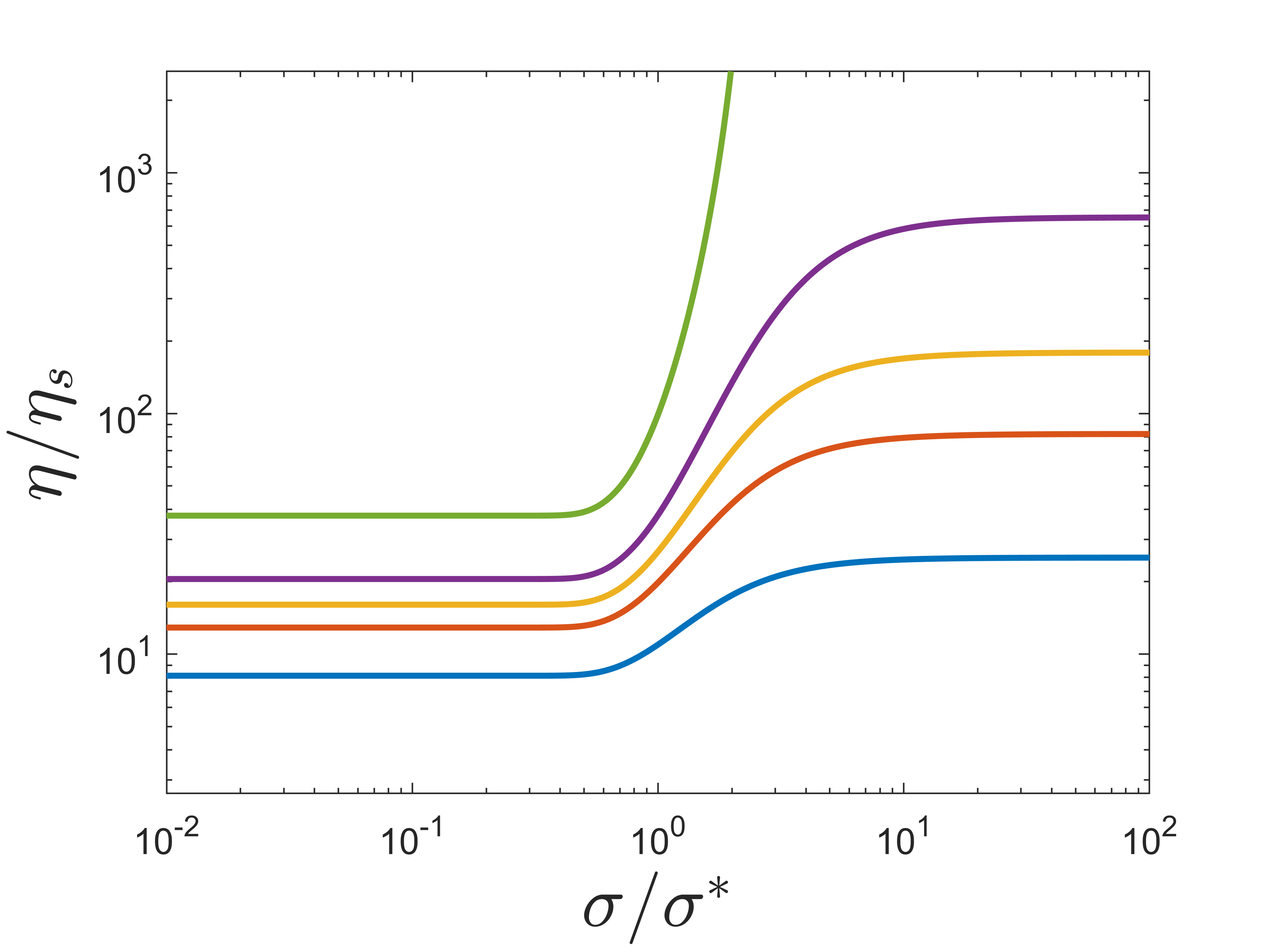}}
\caption{WC model of the DST fluids: 
(a) Flow curves in the stress-shear plane, (b) viscosity dependence on shear stress. Plots for $\sigma^\ast = 0.005$, 
  $\phi_{\mathrm{m}} =0.562$, and $\phi_0=0.693$.}
\label{WC_results}
\end{figure*}

The discontinuous shear-thickening phenomenon wherein the viscosity of the medium undergoes a sharp transition from a lower to a higher value is modeled with the help of a single microstructure scalar field called the fraction of frictional contacts $f(\boldsymbol{x},t)$. 
The use of this scalar field model follows from the Wyarts--Cates model\,\cite{WC2014}, in which the value of $f$ is determined as $\exp(-\sigma^\ast/\sigma)$.
Here, $\sigma^\ast$ is the onset stress at which the suspension begins to shear-thicken. 
Figure\,\ref{WC_results} shows (a) typical flow curves produced by the model and 
(b) the corresponding viscosity dependence on shear stress. 
For sufficiently low volume fractions ($\phi<0.50$), continuous shear-thickening (CST) is achieved. Increasing volume fraction above 0.50 leads to a transition from CST to a DST regime, marked by the appearance of the characteristic S-shape, where stress is a multi-valued function of shear rate. Further increases in volume fraction sharpen the S-curve up to a point of divergence ($\phi=\phi^m$). 

The value of $f(\boldsymbol{x},t)$ at a point in the flow field depends on the local value of stress, and the dependency follows a sigmoid function. 
\citet{Kamrin2019} proposed an evolution equation for $f$ considering the influence of different physical mechanisms towards the formation and destruction of the frictional contacts between suspended particles. 
In the present study, we propose a modified evolution equation combining some features of WC and Baumgarten's model. 
The evolution equation for $f$ is set as  
\begin{gather}
    \frac{df (\boldsymbol{x},t)}{dt}=K_f\dot{\gamma}
    \left(\hat{f}(\boldsymbol{x},t)
    -f(\boldsymbol{x},t)\right)+\alpha\nabla^2f(\boldsymbol{x},t), 
    \label{eq: f evolution}
    \\
    \hat{f}(\boldsymbol{x},t)=
    \exp\biggl(
    -\left[\frac{\sigma^{\ast}}{\sigma_{xy}(\boldsymbol{x},t)}\right]^\beta \biggr),
\end{gather}
where $K_f$ is the microstructural rate constant governing formation and destruction of structure,
and $\dot{\gamma}(\boldsymbol{x},t)
=\sqrt{\dot{\boldsymbol{\gamma}}:\dot{\boldsymbol{\gamma}}/2}$ is the local shear calculated as the second invariant of the strain rate tensor 
$\dot{\boldsymbol{\gamma}}(\boldsymbol{x},t)=\left(\nabla \boldsymbol{v}+\nabla\boldsymbol{v}^{\mathsf{T}}\right)/2$. 
The first term on the right-hand side of \eqref{eq: f evolution} accounts for the local evolution of microstructure, whereas the second term introduces non-local effects via microstructure diffusivity, where $\alpha$ is the diffusion coefficient.

\section{SPH--DST model}
\label{sec_SPH-DST_model}
\subsection{SPH equations of motions}

The Navier--Stokes equations are solved using the SPH methodology. 
A possible discretized form of the governing equations is given below\,\cite{Morris1997}:
\begin{align}
\label{contin_density}
&
\frac{d\rho_i}{dt}= \sum_j m_j \bm{v}_{ij}\cdot\nabla_i W_{ij},
\\ 
&
\begin{multlined}
\label{colagrossi_momentum_eqn}
\frac{d\bm{v}_i}{dt}=
-\sum_{j} m_j \frac{p_i+p_j}{\rho_i \rho_j} 
\nabla_i W_{ij} \\ 
+ 
\sum_{j} \frac{m_j (\eta_i+\eta_j)\bm{x}_{ij}\cdot \nabla_i W_{ij}}{\rho_i\rho_jr_{ij}^2} 
\bm{v}_{ij} + \bm{g}_i,
\end{multlined}
\end{align}
where $m_j$ is the mass of a particle, $W_{ij}=W(\lvert \boldsymbol{x}_{ij} \rvert)$ is the smoothing kernel, $\boldsymbol{v}_{ij}=\boldsymbol{v}_i-\boldsymbol{v}_j$, $\boldsymbol{x}_{ij}=\boldsymbol{x}_i-\boldsymbol{x}_j$,
and $r_{ij}=\sqrt{\boldsymbol{x}_{ij} \cdot\boldsymbol{x}_{ij}}$.

Following the weakly compressible SPH (WCSPH) formulation, a \SI{1}{\percent} variation in density is permitted. The fluid pressure is then calculated based on the Tait's equation of state: 
\begin{equation} \label{eq of state}
p=\frac{\rho_0 c^2}{\gamma} \left[  \left( \frac{\rho}{\rho_{0}}\right)^{\gamma}-1\right] 
+ p_{\mathrm{b}} ,
\end{equation}
where $\rho_0$ is the density of the fluid,
$p_{\mathrm{b}}$ is the background pressure, and $c$ is the numerical speed of sound typically set as $10$ times the maximum flow velocity $v_{\mathrm{max}}$. 


\subsection{SPH--DST microstructure model}

So far, we have described details of SPH for general flows with Navier--Stokes equations. 
In this section, we discuss the closure SPH model for the definition of viscosity $\eta_i(f_i,t)$ in the SPH--DST model.
The constitutive equations are written as follows, where the subscript $i$ indicates the particle index (that is $f(\boldsymbol{x}_i,t)=f_i$):
\begin{gather}
\label{fdot}
\dot{f}_i = K_f\dot{\gamma}_i(\hat{f}_i-f_i)+\alpha\nabla^2f_i,
\\
\label{phi_J}
\phi^{\mathrm{J}}_i = \phi^{\mathrm{m}} f_i +\phi^0 (1-f_i), \\
\label{eta_r_DST}
\eta_i = \eta_{\mathrm{s}} \left(1 - \frac{\phi_i}{\phi^{\mathrm{J}}_i }\right)^{-2}.
\end{gather}
In \eqref{fdot}, $f_i$ is calculated
to evaluate $\phi^{\mathrm{J}}$ in \eqref{phi_J}, 
which in turn is used to obtain viscosity 
via \eqref{eta_r_DST}. 
The value of $\dot{\gamma}$ in \eqref{fdot} is computed as the contraction of the strain rate tensor $\dot{\boldsymbol{\gamma}}$ as follows:
\begin{equation}
\dot{\gamma} = \frac{1}{2}\sqrt{\dot{\boldsymbol{\gamma}}:\dot{\boldsymbol{\gamma}}}, 
\quad
\dot{\gamma}_{\alpha\beta}=\frac{1}{2} \left( \frac{\partial u_{\alpha}}{\partial x_{\beta}}+\frac{\partial u_{\beta}}{\partial x_{\alpha}} \right).
\label{eq:shear_rate}
\end{equation}
The SPH approximation of the gradient of the $\alpha$-component of the velocity vector in $\beta$-direction is obtained as
\begin{equation}
\left( \frac{\partial u^\alpha}{\partial x^\beta}\right) _i = \sum_j \frac{m_j}{\rho_j}(u^\alpha_j-u^\alpha_i) \nabla_i^\beta W_{ij}.
\end{equation}
The non-local term in equation (11) is discretised in a manner consistent with the Morris formulation of the moment equation\,\cite{Lind2020},
\begin{equation}
    (\nabla^2f)_{i\in f}=\sum_{j\in f}\frac{2m_j(f_i-f_j)\boldsymbol{x}_{ij}\cdot\nabla_iW_{ij}}{\rho_jr^2_{ij}}.
\end{equation}
To simulate the `S' - curve 
(see Fig.\,\ref{WC_results}) with the above definition of order parameter $f$, the choice of $\phi$ has to be sufficiently close to $\phi_{\mathrm{m}}$ which in the above case is $\phi>0.50$. 
However, the value of $\eta/\eta_{\mathrm{s}}$ 
for $\phi=0.54$ at a high shear rate is approximately 3000, which drastically lowers the time step size required for stable integration of the governing equations. 
A simple idea used to address this was to modify the definition of $f$ so that we can anticipate the `S' at a lower volume fraction. 
\begin{figure*}[tb]
\subfloat[][]{\includegraphics[width=0.4\textwidth]{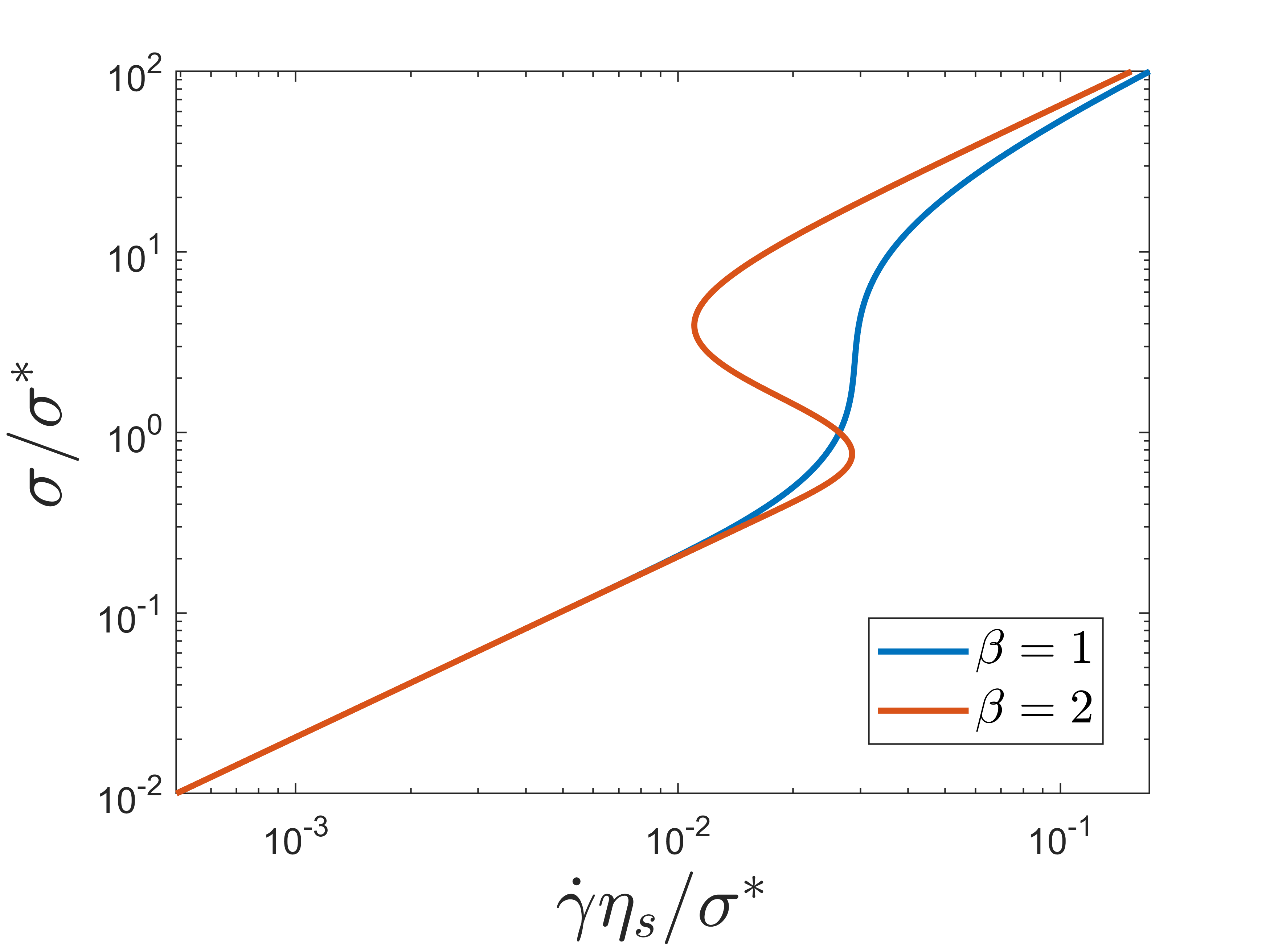}}
\subfloat[][]{\includegraphics[width=0.4\textwidth]{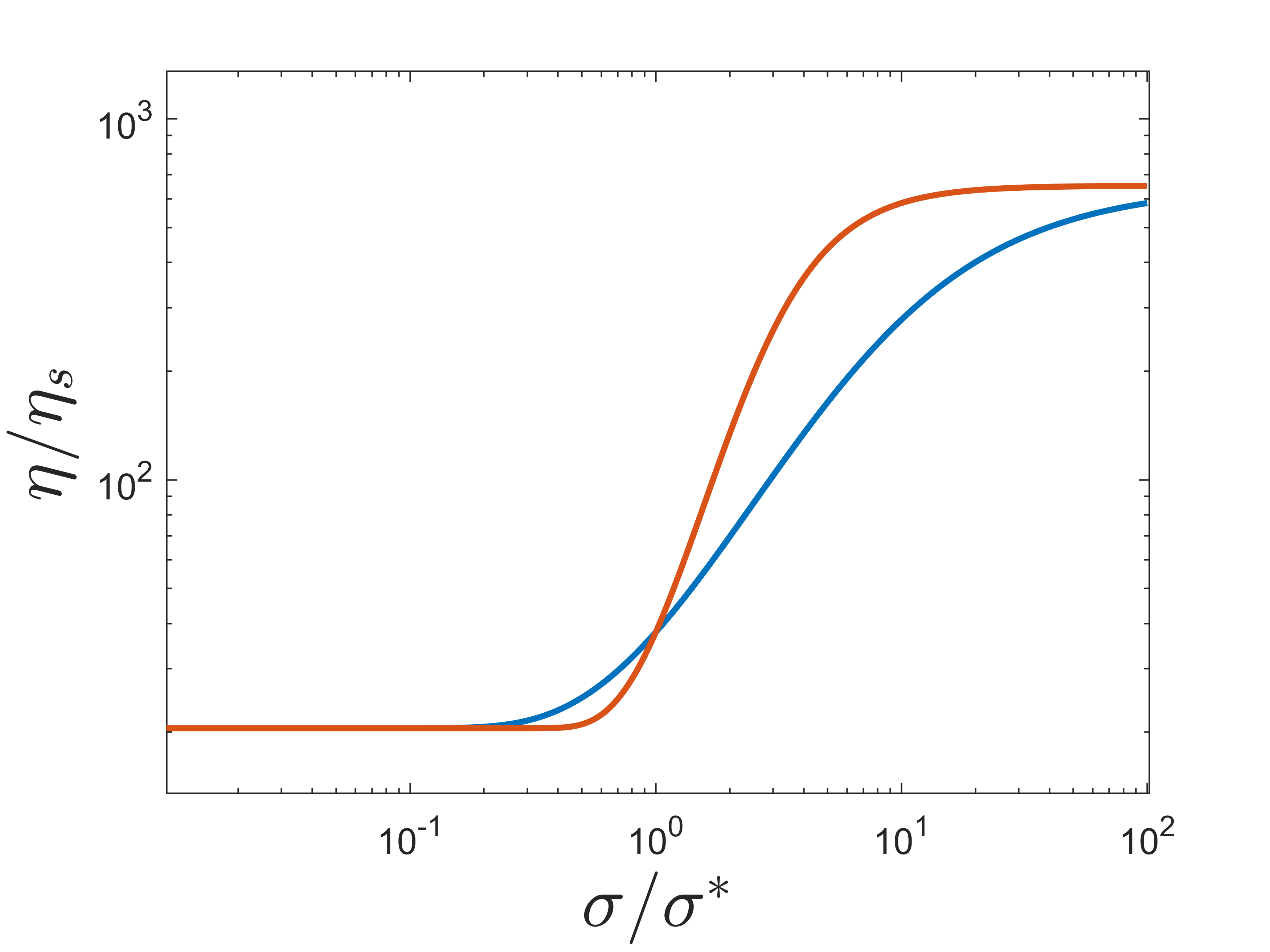}}
\caption{Comparison of WC model flow curves, 
i.e., (a) rate dependence of viscosity and (b) stress dependence of viscosity for $\beta=1$ and $\beta=2$. \label{mod_f}}
\end{figure*}
Therefore, the modified definition of $f$ is as follows.
\begin{equation}
\hat{f}_i = \exp\bigl\{-(\sigma^{\ast}/\sigma_i)^\beta\bigr\},
\end{equation}
where $\beta$ is the parameter setting the `sharpness' of the transition between the frictional and frictionless plateaus, without changing the plateau values. We set $\beta=2$ in this work.
Increasing it allows us to introduce the previously discussed S-shaped rheology to an otherwise CST volume fraction, resulting in DST behaviour without increasing viscosity to computationally prohibitive values. The rest of the parameters are obtained by fitting the model to experimental data and other models.
Once the value of $f_i$ and $\phi^{\mathrm{J}} (f_i)$ are obtained, the dynamic viscosity of a SPH particle is updated using \eqref{eta_r_DST}.

\subsection{Solid wall boundary modeling}

The dummy particle method of \citet{Adami2012} is employed for imposing no-slip velocity boundary conditions and impermeability conditions. 
Pressure, density, and velocity are assigned to the dummy particles as
\begin{gather}
  p_{i\in w} = \frac{\sum\limits_{j\in f} p_{j} W_{ij} + (\bm{g}-\bm{a}_{i\in w}) 
  \cdot \sum\limits_{j\in f}\rho_{j} \bm{x}_{ij} W_{ij}} {\sum\limits_{j\in f} W_{ij}}, \\
 \bm{a}_{i\in w} =-\frac{\nabla p_{i\in f}}{\rho_{i\in f}}+\bm{g}, \\
\rho_{i\in w} = \rho_{0} \left(\frac{p_{i\in w} - p_{\mathrm{b}}}{p_0} + 1 \right)^\frac{1}{\gamma}, \\
p_0=\frac{\rho_0 c^2}{\gamma}, \\
\bm{v}_{i\in w} = 2\bm{v}_{\mathrm{wall}} - \tilde{\bm{v}}_{i\in w},
\end{gather}
where subscript \textit{w} and \textit{f} 
denote the dummy wall and fluid particles, respectively,
$\bm{v}_{i\in w}$ is the prescribed wall velocity, and
\begin{equation}
\tilde{\bm{v}}_{i\in w} = \frac{\sum\limits_{j\in f}{\bm{v}_j W_{ij}}}{\sum\limits_{j\in f} {W_{ij}}}.
\end{equation}

\subsection{Time integrators}
A semi-implicit predictor-corrector type integration scheme is used for time marching as follows.

\noindent
(1) Predictor step:
\begin{gather}
\bm{v}_i^{n+\frac{1}{2}} = \bm{v}_i^{n} + \frac{\Delta t}{2}\left(\dfrac{d\bm{v}_i}{dt}\right)^n,\\
\rho_i^{n+\frac{1}{2}} = \rho_i^{n} + \frac{\Delta t}{2} \left(\dfrac{d\rho_i}{dt} \right)^{n}, \\
\bm{x}_i^{n+\frac{1}{2}} = \bm{x}_i^{n} + \frac{\Delta t}{2} \left(\bm{v}_i\right)^n, \\
\label{fdot_2}
f^{n+\frac{1}{2}}_i = f^{n}_i + \frac{\Delta t}{2}\left(  \frac{df_i}{dt} \right)^n. 
\end{gather}

\noindent
(2) Corrector step:
\begin{gather}
\bm{v}_i^{n+\frac{1}{2}} = \bm{v}_i^{n} + \frac{\Delta t}{2}\left(\dfrac{d\bm{v}_i}{dt}\right)^{n+\frac{1}{2}}, \\
\rho_i^{n+\frac{1}{2}} = \rho_i^{n} + \frac{\Delta t}{2} \left(\dfrac{d\rho_i}{dt} \right)^{n+\frac{1}{2}},\\ 
\bm{x}_i^{n+\frac{1}{2}} = \bm{x}_i^{n} + \frac{\Delta t}{2} \left(\bm{v}_i\right)^{n+\frac{1}{2}},\\ 
\label{fdot_2}
f^{n+\frac{1}{2}}_i = f^{n}_i + \frac{\Delta t}{2}\left(  \frac{df_i}{dt} \right)^{n+\frac{1}{2}}.
\end{gather}

The flow variables at the subsequent time step are obtained as follows:
\begin{gather}
\bm{v}_i^{n+1} = 2\bm{v}_i^{n+\frac{1}{2}} - \bm{v}_i^n, \\
\rho_i^{n+1} = 2\rho_i^{n+\frac{1}{2}} - \rho_i^n, \\
\bm{x}_i^{n+1} = 2\bm{x}_i^{n+\frac{1}{2}} - \bm{x}_i^n, \\
\label{fdot_2}
f^{n+1}_i = 2f^{n+\frac{1}{2}}_i - f^{n}_i, 
\end{gather}
where $n$ represents the current time instant, $n+ \frac{1}{2}$ represents the predicted time instant (variables at an intermediate time level), and $n+1$ represents the corrected time instant. 
This scheme is second order accurate in time. Finally, the time-step size $\Delta t$ is determined based on the following condition
\begin{equation}
\Delta t = \min \left\lbrace {0.25 \frac{h_{\mathrm{sl}}}{c}, 0.125 \frac{h_{\mathrm{sl}}^2}{\nu}, 0.25 \left(\frac{h_{\mathrm{sl}}}{|\bm{g}|}\right)^{1/2}}\right\rbrace ,
\end{equation}
where $h_{\mathrm{sl}}$ is the smoothing length, and $\nu$ is the kinematic viscosity.

\section{Numerical Simulations}
\label{sec_Numerical_Simulations}

\subsection{Stress-Imposed Shear Flow}
\label{subsec_Stress-Imposed_Shear_Flow}

The behaviour of the previously introduced DST model is explored by simulating the rheology in a simple shear geometry. The fluid is bound between two solid walls vertically and two periodic boundaries horizontally at a distance $l=\SI{0.01}{m}$.
The domain is planar with no depth in the vorticity direction. Unlike previous shear-imposed simulations\,\cite{JFM2019}, input stress $\sigma_{\mathrm{in}}$ is set here on the upper wall by assigning wall particles an appropriate velocity.

\begin{figure*}
    \centering
     
    \subfloat[][]{\includegraphics[width=0.3\textwidth]{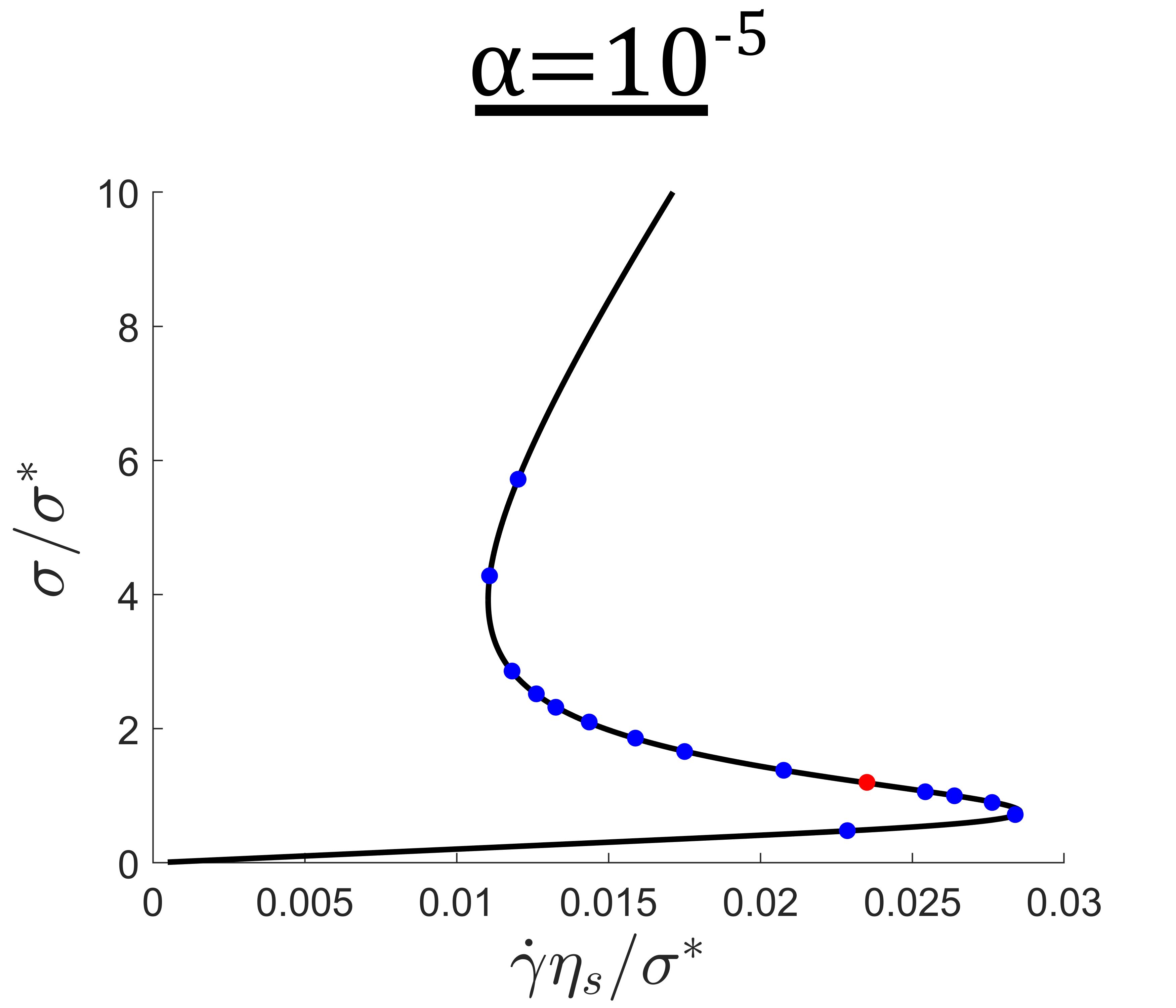}}
    \subfloat[][]{\includegraphics[width=0.3\textwidth]{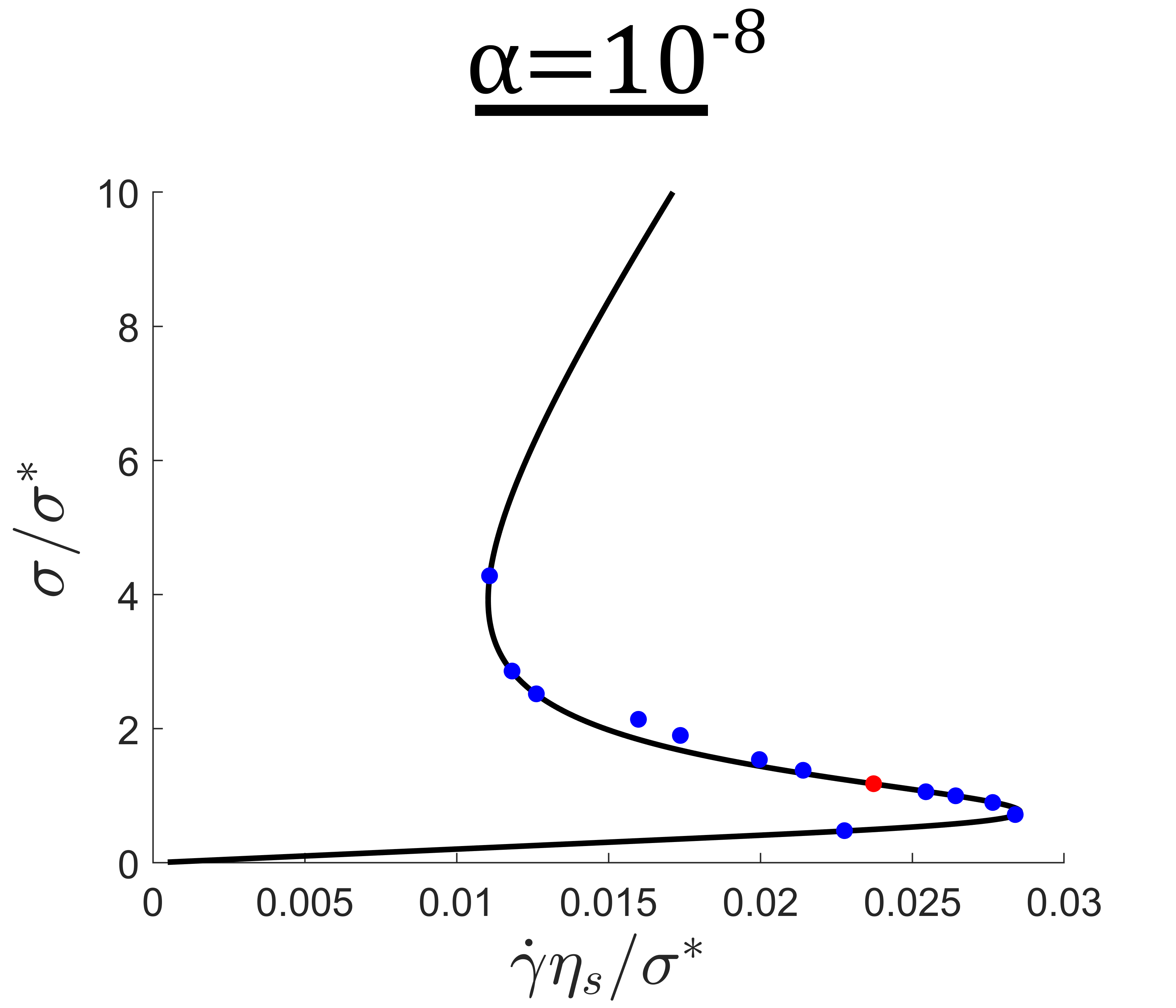}}
    \subfloat[][]{\includegraphics[width=0.3\textwidth]{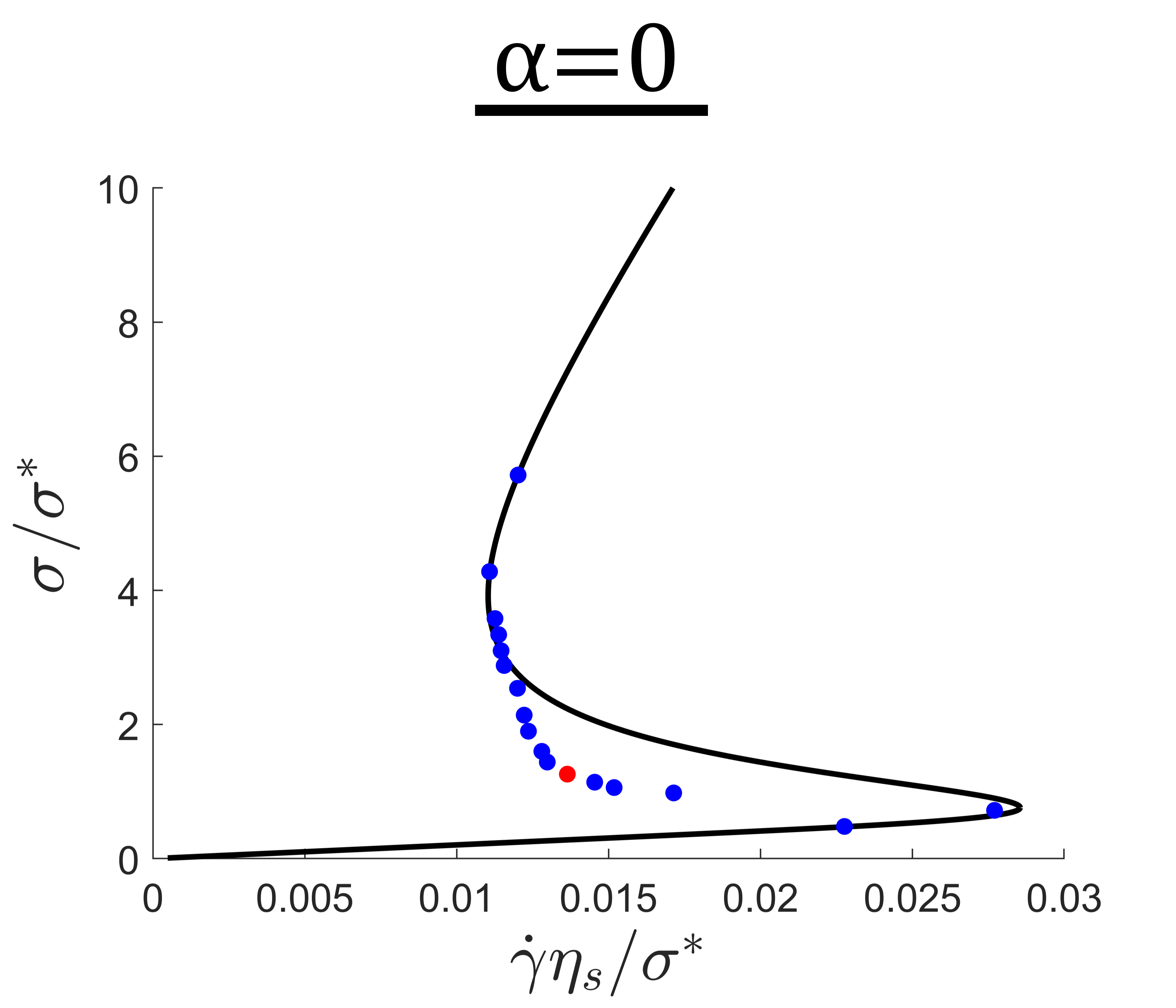}}\\
    \subfloat[][]{\includegraphics[width=0.3\textwidth]{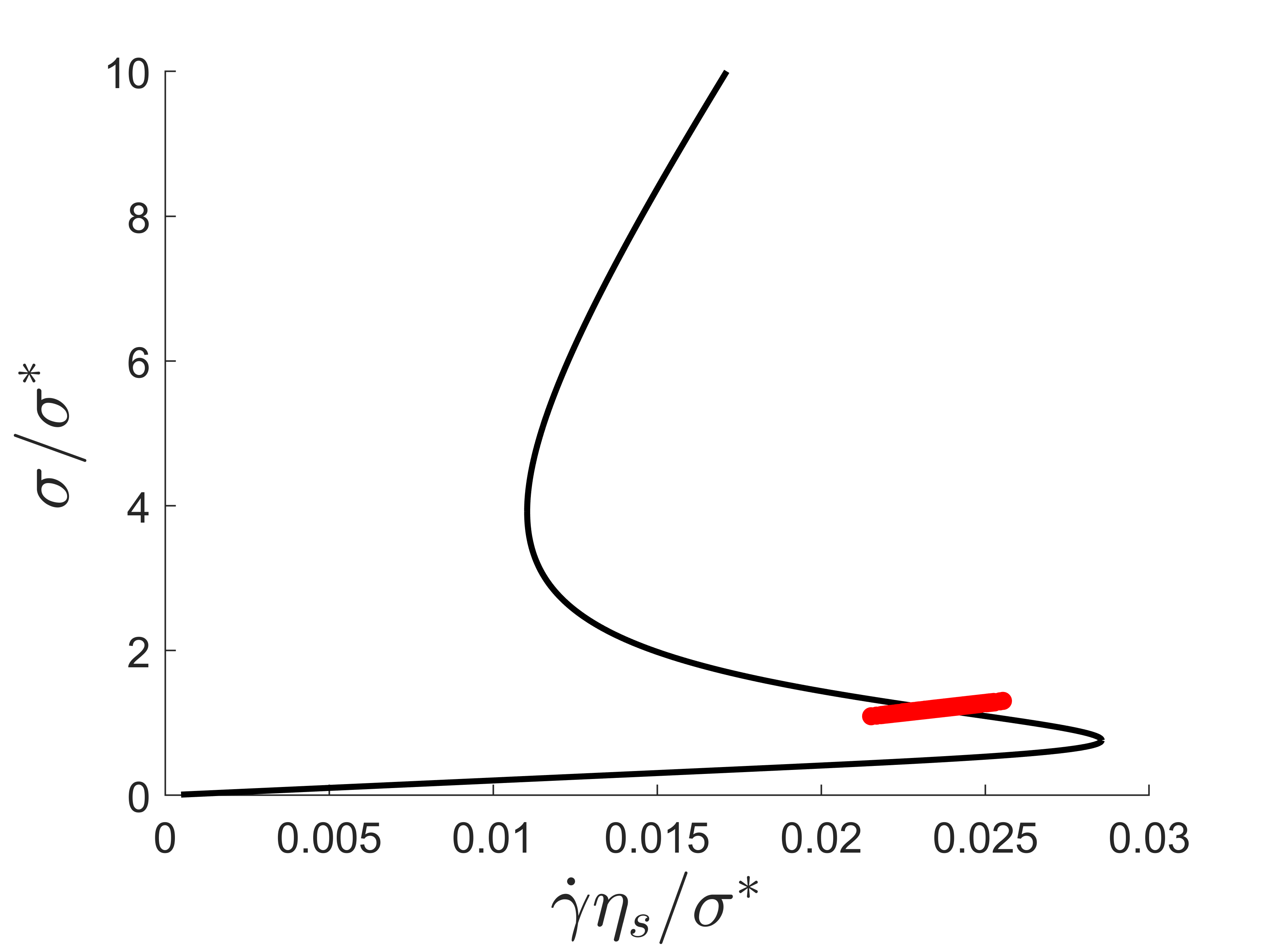}}
    \subfloat[][]{\includegraphics[width=0.3\textwidth]{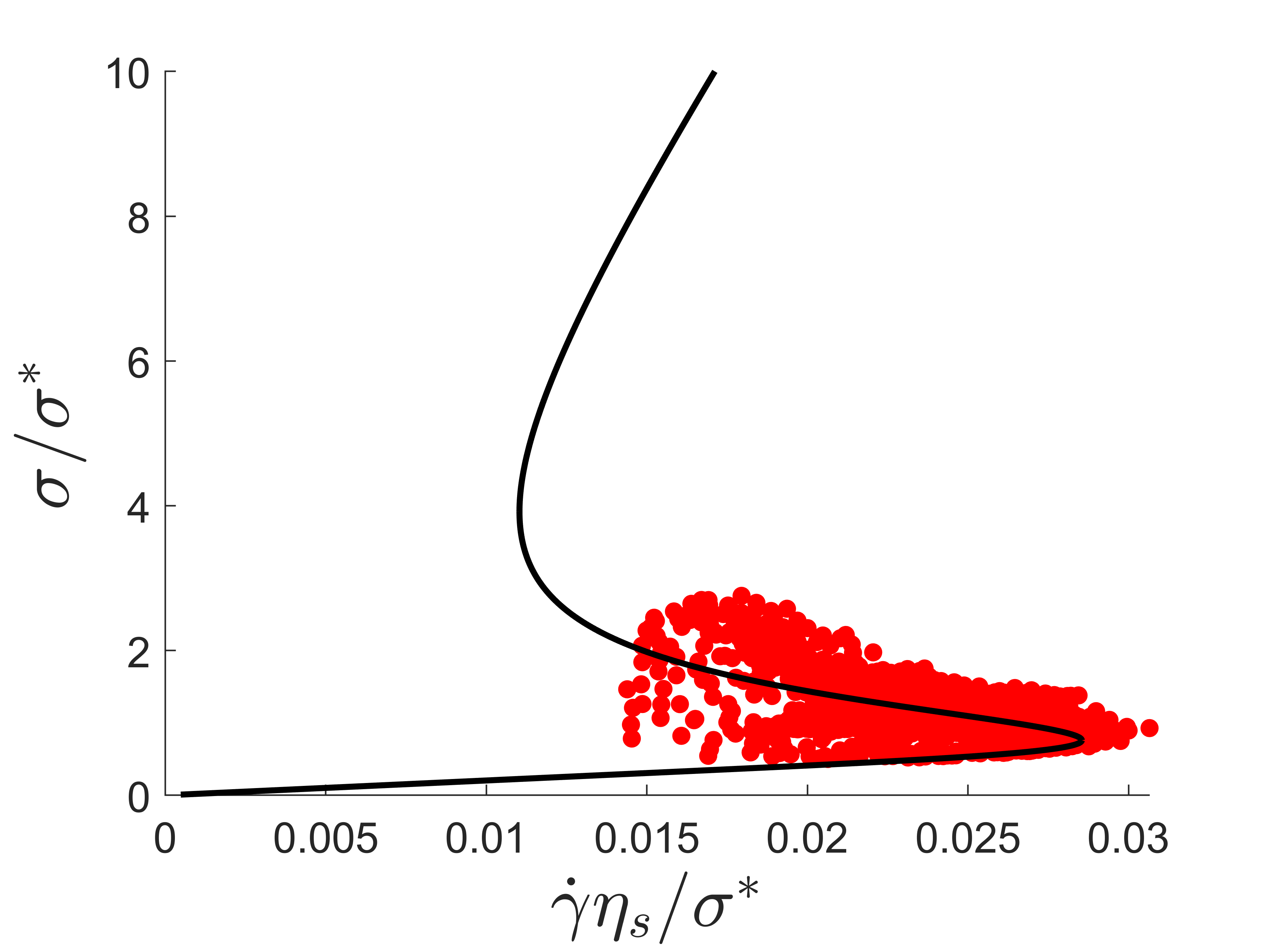}}
    \subfloat[][]{\includegraphics[width=0.3\textwidth]{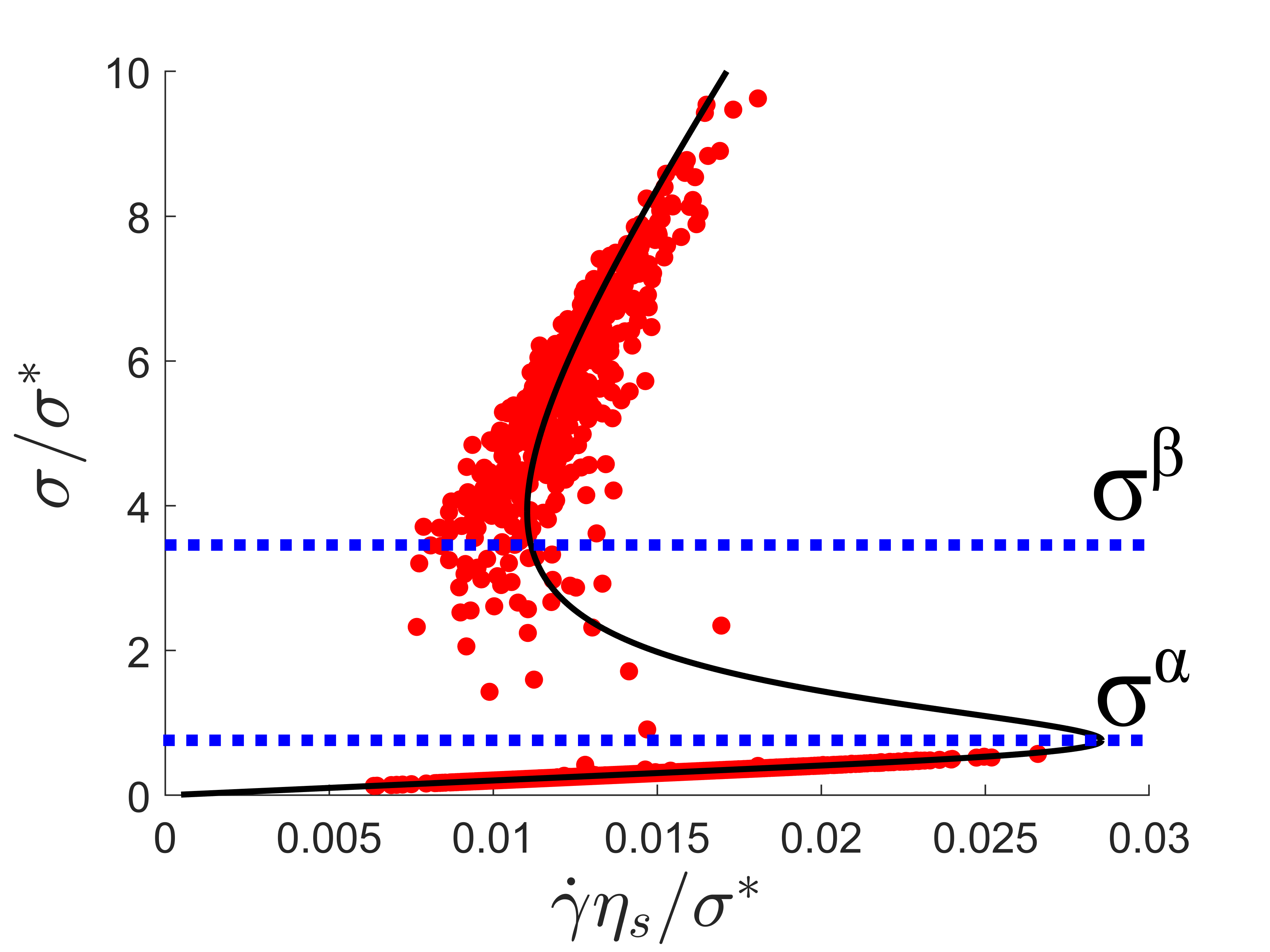}}\\
    \quad\qquad
    \subfloat[][]{\includegraphics[width=0.3\textwidth]{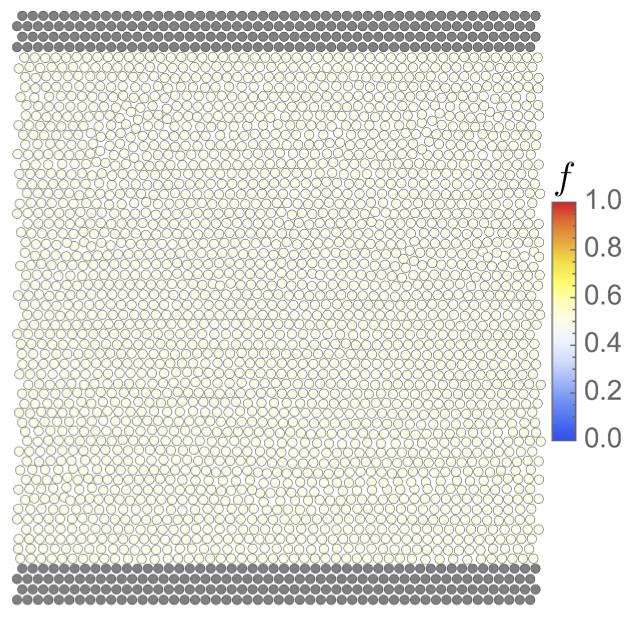}}
    \subfloat[][]{\includegraphics[width=0.3\textwidth]{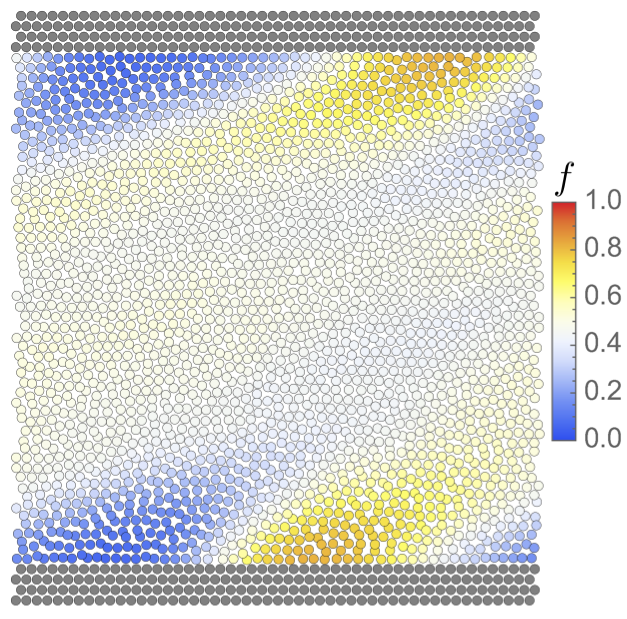}}
    \subfloat[][]{\includegraphics[width=0.3\textwidth]{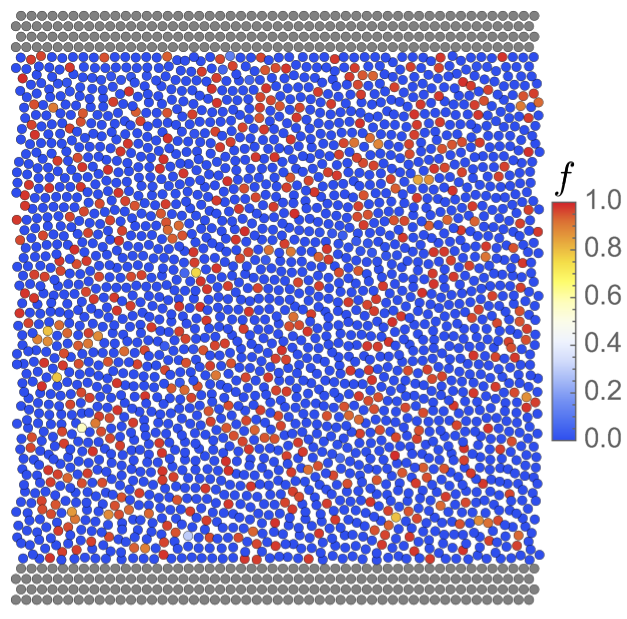}}

    \caption{%
      The results of the SPH simulation compared against the WC model (black) for strong ((a), (d), and (g)), moderate ((b), (e), and (h)), and weak ((c), (f), and (i)) non-locality. 
      The top row (a)--(c) consists of domain-time averaged shear stress and shear rate signal (SPH in blue and red). A single simulation (red) in each diagram ($\Sigma_E=1.25$ was chosen, and values of all SPH particles were plotted over the WC flow curve (d)--(f). Fields of the microstructure parameter $f$ visualised in (g)--(i).}
    \label{WCmodel_vs_Sim}
\end{figure*}

To impose specified stress on the top wall of the Couette geometry, the following approach is used. 
The wall particles are assigned with velocity ($v_{i\in w}$) computed from the wall force 
($F_{\mathrm{diff}}$), which is in turn determined from a predictor--corrector method. 
\begin{equation}
v_{i\in w}^{n+1} = 
v_{i\in w}^n + 
K_p F_{\mathrm{diff}}\Delta t. 
\label{eq:velocity_increment}
\end{equation}
The net wall force 
($F_{\mathrm{diff}}$) is computed as the difference between the applied wall force 
($F_{\mathrm{in}}$) and the resistive shear force ($F_{\mathrm{r}} $) exerted by the fluid on the top wall 
\begin{gather}
F_{\mathrm{diff}} = F_{\mathrm{in}} - F_{\mathrm{r}},  \\
F_{\mathrm{r}} = \sum_{j\in w}F_{j}, \\
F_{i\in w} = 
\sum_{j\in f} F_{ij}^{\nu},
\label{eq:F_i}
\end{gather}
where subscripts w and f denote wall and fluid particles, respectively.
$F_{ij}^{\nu}$ is the inter-particle viscous force in eq.\,\eqref{colagrossi_momentum_eqn}.
\begin{table*}[tb]
\caption{\label{tab:table1}Simulation Parameters}
\begin{ruledtabular}
\begin{tabular}{c|c|c|c|c|c|c|c}
$h$\,[\si{\metre}] & 
$l$\,[\si{\metre}] &
$\rho$\,[\si{\kilogram . \metre^{-3}}] & 
$\eta_s$\,[\si{\pascal \second}] & $\phi$ & 
$\phi^m$ & 
$\phi^0$ & 
$\sigma^{\ast} \, [\si{\pascal}]$ \\
\hline
0.01 & 0.01 & 1000 &
0.001 &
0.48/0.50/0.54& 
0.562 & 0.693 & 0.005
\end{tabular}
\end{ruledtabular}
\end{table*}
All simulations use a quintic kernel, 
and smoothing length $h_{\mathrm{sl}}=1.1\Delta x$, where $\Delta x=0.02l$ is the initial spatial resolution.
Simulations are carried out for the same material parameters (see table\,\ref{tab:table1}) with different imposed stress values. 
Three microstructure diffusion coefficients are tested to probe effects of the non-local term --- strong non-locality ($\alpha=\SI{e-5}{\metre^2 . \second^{-1}}$), moderate non-locality ($\alpha=\SI{e-8}{\metre^2 . \second^{-1}}$), 
and weak non-locality ($\alpha= \SI{0}{\metre^2 . \second^{-1}}$). 
In addition, three volume fractions were simulated for the case of strong non-locality: 
CST ($\phi=0.48$), moderate DST ($\phi=0.50$), and strong DST ($\phi=0.54$). 
Steady-state values are used to construct the flow curve (Fig.\,\ref{WCmodel_vs_Sim}). 
All rheometric values are obtained by averaging over the entire domain and across time, 
with an equilibration time of \SI{100}{\second}.
Resolution dependence of the microstructure was probed, and $50 \times 50$ particle configuration was found sufficient.

Far below onset stress, behaviour is essentially Newtonian --- all three measured properties (shear stress, shear rate, and microstructural parameter $f$) approach their respective steady-state values asymptotically and within relatively short timescales. 
Once the stress is increased, the time required to achieve a steady state is significantly increased, and over/under-shoots are present prior to achieving a steady state. 
This is likely due to the formation of microstructure being relatively slow compared to purely inertial transient effects --- in the initial stages, the lack of structure allows shear rate to peak above the equilibrium value. 
This overshoot then accelerates formation of microstructure, increasing viscosity, which in turn adjusts the shear rate (or undershoots).
To mitigate this behaviour, we choose to initialise all simulations from a pre-defined homogeneous field $f$ corresponding to the correct prediction of the WC at a given input stress, along with the appropriate shear rate via the upper wall velocity.
\begin{figure}
    \centering
    \includegraphics[width=0.45\textwidth]{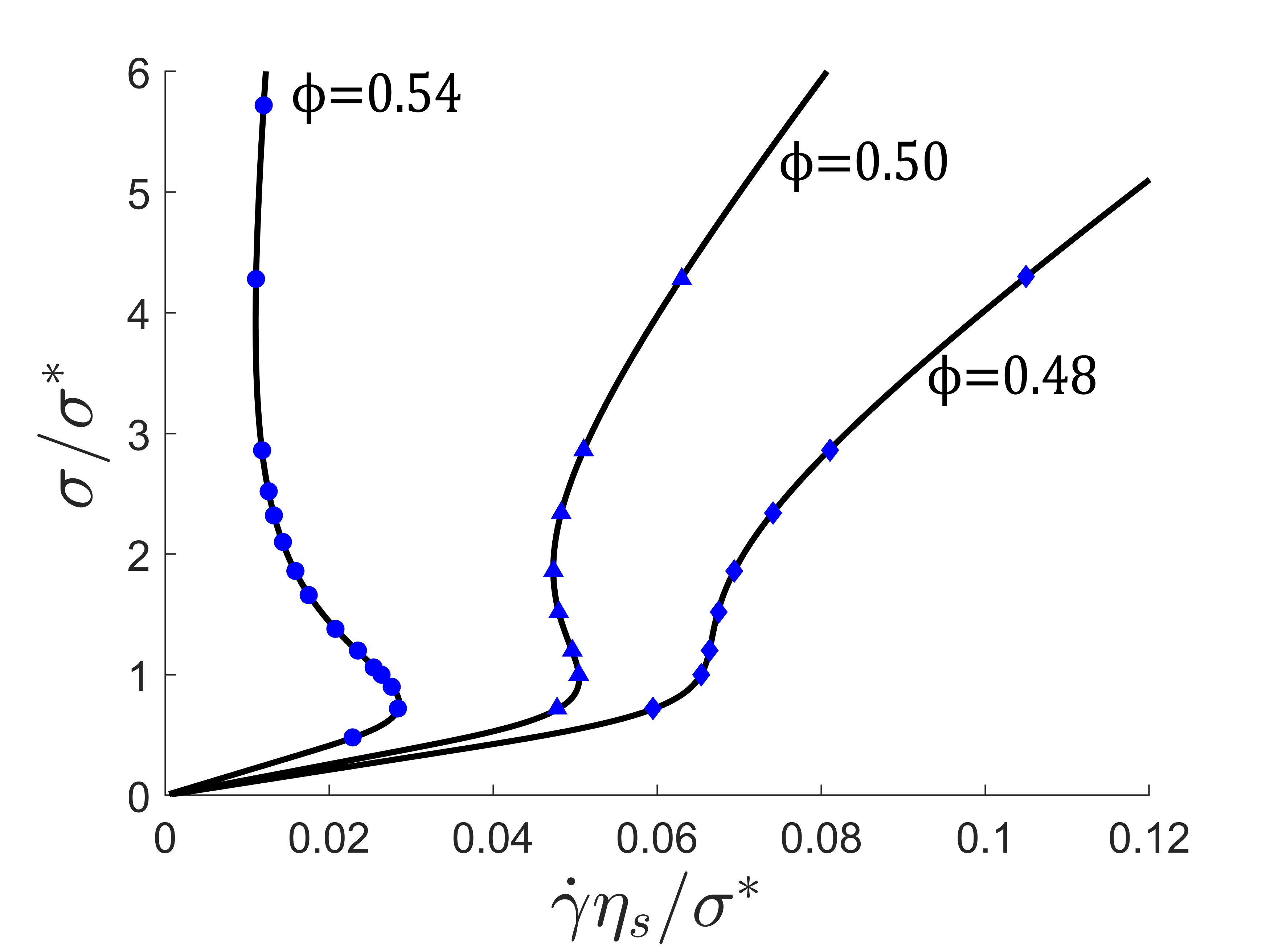}
    \caption{Results for three volume fractions: 0.54 (circles), 0.50 (triangles), and 0.48 (diamonds). Strong non-locality ($\alpha=10^{-5}$). }
    \label{all phi}
\end{figure}
For all values of $\alpha$, simulation results (Fig.\,\ref{WCmodel_vs_Sim}(a)--(c)) on stable branches are in excellent agreement with the theoretical model predictions. 

In the case of strong non-locality, the simulation results for all three volume fractions (Fig.\,\ref{all phi}(a)) match the WC model very well across the entire flow curve, including the region of negative gradient for the DST volume fractions. 
Both the stress and microstructure fields are entirely homogeneous Fig.\,\ref{WCmodel_vs_Sim}(d) and (g) due to the strong non-local term, with minor deviations in the computed shear rate field, likely due to the imprecision associated with numerical approximations and low resolution.


In the purely local ($\alpha=0$) case, for all imposed stresses falling within the unstable region of negative flow curve gradient, seemingly spurious `jumps' in microstructure, local stress and viscosity were observed, where individual particles attain either much higher or lower stress and microstructural parameter with respect to the imposed stress. 
This behaviour had no impact on the numerical stability of the simulations, and a solution could be computed even in this extreme case. 
These jumps occurred with no clear spatial correlation, yet both the domain averaged properties, and the properties measured at the wall remained in close agreement with the correct theoretically predicted values. 
By plotting all individual SPH particles for a given applied stress over the expected flow diagram (Fig.\,\ref{WCmodel_vs_Sim}(f)), it is apparent that these jumps are directed towards the valid stable solutions of the WC model.

In principle, a DST fluid could undergo vorticity banding under stress-controlled regime, which, in the simplest case, would manifest as any stress applied within the unstable branch (the branch joining frictional and frictionless branches) being split vertically into two bands organised along the vorticity direction each with either higher or lower stress, but the same shear rate. The width of the bands is then decided by a lever rule, such that the total average stress is equal to the applied stress. 
Such phenomena in DST materials have been experimentally observed in the work of \citet{Herle}, but \citet{Hermes2016} excludes the possibility of steady-state vorticity bands in non-Brownian suspensions on the account of particle migration and normal stresses. 
For a purely local model and a constant volume fraction field, no effects counteract such splitting, giving rise to a new steady state configuration, where all individual SPH particles occupy stable branches of the flow curve ($\frac{df_i}{dt}=0$), while being split such that the stress control equation is also satisfied ($\frac{dv_{\mathrm{wall}}}{dt}=0$).

%

The lack of spatial correlation in structure (Fig.\,\ref{WCmodel_vs_Sim}(i)) is caused by the action of the microstructure evolution equation on the numerical noise introduced in the SPH solution. 
Suppose all particles are initialised exactly at the steady state solution, with small perturbation in a random direction in the rate-stress plane. 
The particles that find themselves to the right of the flow curve will experience $\frac{df_i}{dt}>0$, leading to an increase in local viscosity. 
This will, in turn, increase the local shear stress, increasing the value of $\hat{f_i}$, ensuring that the particle will continue to move upwards until it reaches a stable upper branch where $\frac{df_i}{dt}=0$. 
Identical argument applies to the particles to the left of the curve, with the resultant downward movement instead. 
In the case of our simulations, the small perturbations are introduced by the inherent numerical noise associated with meshless calculation the local shear rate $\dot{\gamma}_i$ via eq.\,\eqref{eq:shear_rate}.
These small imprecisions do not have a spatial correlation, hence after stress-splitting amplification we arrive at effectively random spatial distribution of microstructural parameter. 
The left-right split of particles should approach an even split ($\psi^1=\psi^2=0.5$); however, the flow curve is not symmetric about the horizontal imposed stress line, and such split would result in an increase of measured shear stress. 
This means that as the stress-splitting takes place, particles are simultaneously rearranged between the lower and upper branches via the stress-control mechanism of 
eqs.\,\eqref{eq:velocity_increment}--\eqref{eq:F_i} to achieve correct average stress.


To further explore both the microstructural transients and steady states, new dimensionless parameters are defined:
\begin{gather}
    \psi^1=\frac{\text{number on the lower branch}}{\text{total number}},
\\
    \psi^2=\frac{\text{number on the upper branch}}{\text{total number}},
\\
    \psi^{\mathrm{u}}=\frac{\text{number on the unstable branch}}{\text{total number}}.
\end{gather}
The number of particles on either the upper or lower branch is measured with reference to the two threshold stress values $\sigma^{\alpha}$ and $\sigma^{\beta}$, which occur at the intersection of the unstable branch with the upper and lower branches, respectively.  
Typical response consists of a \SI{50}{\second} period for banding process to begin and almost \SIrange{150}{200}{\second} to reach a steady state.
Different initial configurations were tested including a) relaxed microstructure, b) correct steady-state microstructure, and c) excessive level of structure. 
Due to the local nature of this process, the splitting is independent of the resolution as there is no finite characteristic length scale associated with this process.
All three configurations yielded the same steady-state banding 
ratios, i.e., the above-mentioned steady-state is broadly independent of the path taken to achieve it. 
Furthermore, shear-controlled cases were tested both within and outside the multivariate region and yielded results with no apparent banding; a result consistent with the theory where only stress-controlled flow of DST fluids should experience vorticity banding.
\begin{figure}[h]
\centering
\includegraphics[scale=0.45]{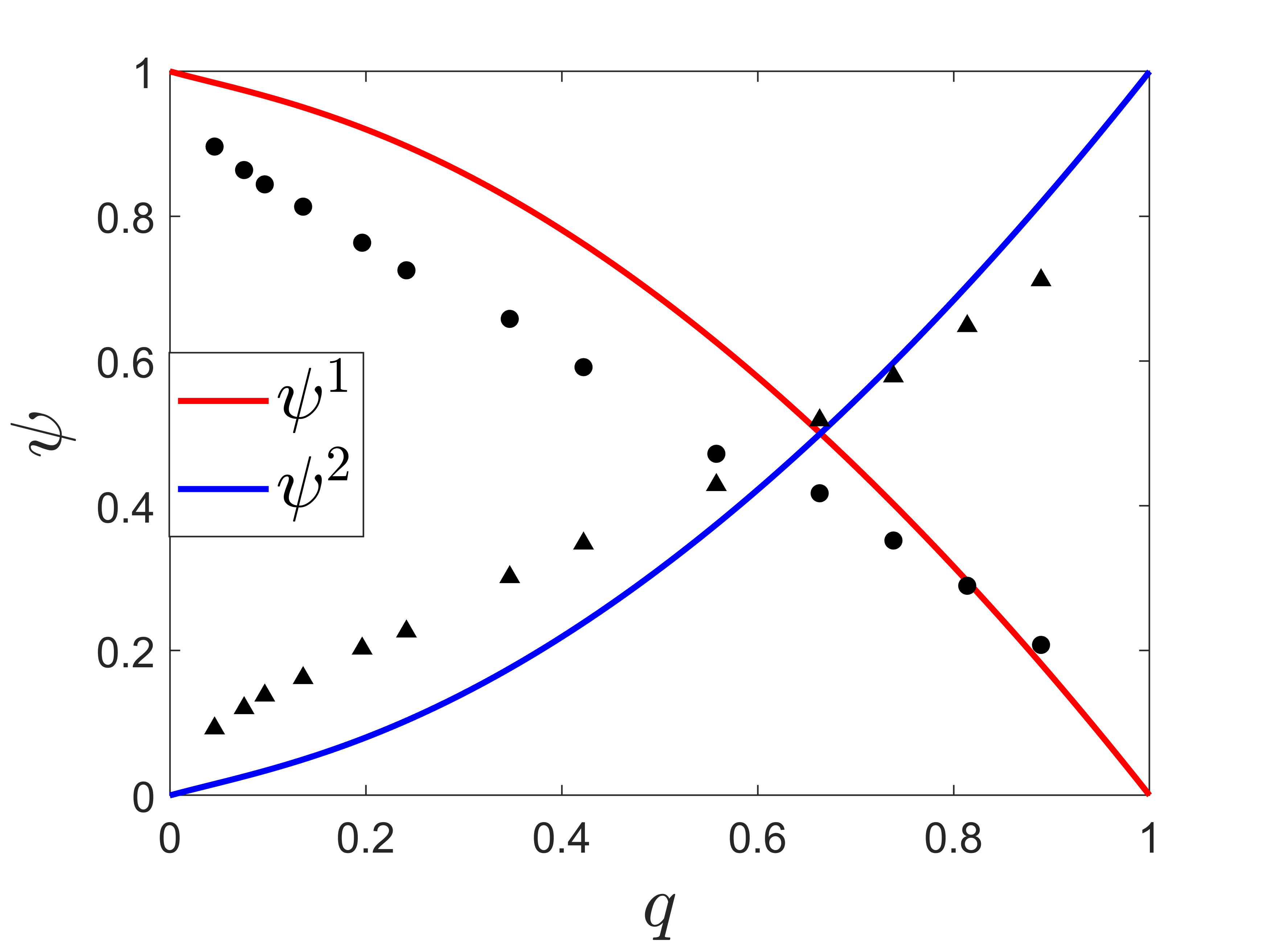}
\caption{\label{branch_split_ratios}
Theoretically predicted branch split ratios (lines) vs.~simulations. Simulation values represented by black circles ($\psi^1$) and triangles ($\psi^2$) .}
\end{figure}

To better understand the exact split between $\psi^1$ and $\psi^2$ at steady-state, a simple model of DST vorticity banding is considered.
Suppose that stress is applied within the unstable region of the flow curve at a coordinate $(\dot{\gamma}^{\mathrm{u}}, \sigma^{\mathrm{u}})$. 
This unstable point will split directly in the vertical direction until it meets the lower and upper branches, resulting in a new steady-state consisting of two stable points $(\dot{\gamma}^{\mathrm{u}},\sigma^1)$ 
and $(\dot{\gamma}^{\mathrm{u}},\sigma^2)$, respectively. We also assume that the total average stress is preserved such that:
\begin{equation}
\sigma^{\mathrm{u}} = \psi^1\sigma^1+\psi^2\sigma^2.
\label{lever rule}
\end{equation}
By considering that the system is now in a steady state
($ \psi^{1}+\psi^{2}=1$ and $\psi^{\mathrm{u}}=0$),
both fractions can be uniquely determined for any input stress. 
It is immediately apparent that the exact split will be governed by the exact shape of the flow curve. 
Finally, we define a dimensionless stress coordinate $q$ using critical stress values $\sigma^{\beta}$ and $\sigma^{\alpha}$ (see Fig.\,\ref{WCmodel_vs_Sim}(f)):
\begin{equation}
q=\frac{\sigma^{\mathrm{u}}-\sigma^{\beta}}{\sigma^{\alpha}-\sigma^{\beta}},
\end{equation}
such that, for $q<0$, input stress is on the lower stable branch, 
$0\leq q\leq 1$
input stress is on the unstable branch, and for $q>1$, input stress is on the upper stable branch. 
The results along with simulations are presented in Fig.\,\ref{branch_split_ratios}. 
The simulation results exhibit a significantly lower number of particles on the lower branch and a higher number of particles on the upper branch relative to the theoretical predictions.
This is due to how the split rearrangement is achieved. Since an even split would result in an increase in the measured shear stress, initial splitting causes the stress-control scheme to reduce shear rate. The shear rate reduction is set by the decrease in wall velocity, which affects all particles, causing a shift to the left in the ($\dot{\gamma}$--$\sigma$) phase diagram.
 Particles on the upper branch can only shift to the lower branch by crossing the critical stress $\sigma^{\beta}$, where particles can no longer follow the linear relationship between shear rate and shear stress, and 'fall' down to the stable branch, due to the combined effects of the microstructure evolution equation and the stress-control scheme.
 This results in the particles on the upper stable branch, occupying shear stress and shear rate values lower than anticipated in the naive vertical splitting scenario. 
 Lower stress value of $\sigma^1$ in eq.\,\eqref{lever rule} necessitates higher value of $\psi^2$ (and lower value of $\psi^1$) at steady state. 
 This mechanism also accounts for the poor agreement between the simulations and the WC model Fig.\,\ref{WCmodel_vs_Sim}(c) --- average shear stress is close to the correct input value, but the splitting causes much lower shear rate values.

 Finally, the case of moderate non-locality is considered. The averaged results show good agreement with the WC model Fig.\,\ref{WCmodel_vs_Sim}(b) along the entire flow curve, just as in the homogeneous case. 
 Unlike the homogeneous case, the plot of individual SPH particles in Fig.\,\ref{WCmodel_vs_Sim}(e) shows a clear and significant spread in both shear stress and shear rate. However, this spread is different 
 from the local case --- particles occupy mainly the unstable region between the two stable branches. 
 The bands in the microstructure parameter are also accompanied by bands in shear stress and shear rate.
 This steady state occurs as a result of a balance between the split-inducing local term and the smoothing non-local term in the microstructure evolution equation. Furthermore, the competition between the local and non-local effects yields an apparent spatial correlation in the microstructure field Fig.\,\ref{WCmodel_vs_Sim}(h) reminiscent of the bands observed in the work of \citet{Nakanishi}.

\subsection{Channel Flow}
\label{subsec_Channel_flow}

\begin{figure}[hbt]
    \centering
    \includegraphics[width=0.45\textwidth]{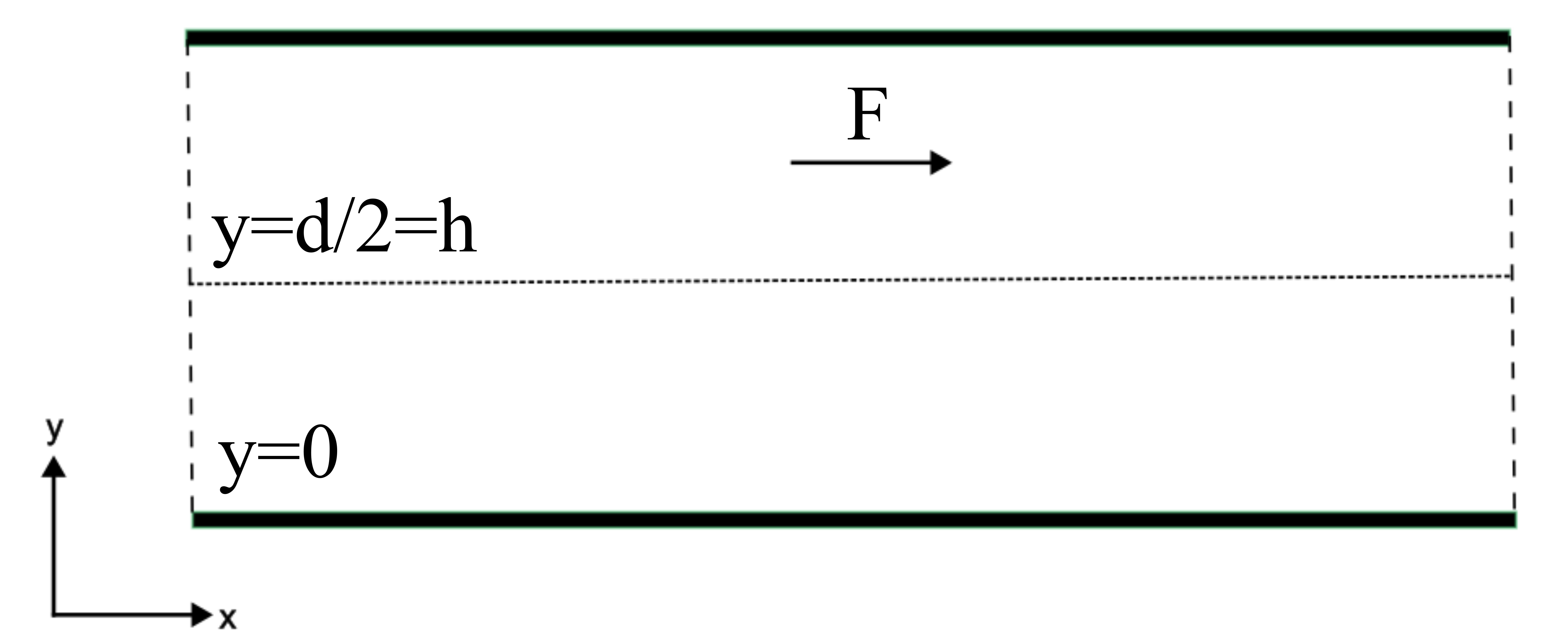}
    \caption{Channel geometry sketch}
    
    \label{channel_geometry}
\end{figure}

\subsubsection{Theoretical Analysis}
In this section, we consider flow in a planar channel 
(Fig.\,\ref{channel_geometry}), with gap $d=\SI{0.01}{m}$
and periodic boundary conditions in the flow direction. 
Body force $F$ parallel to the flow direction is applied to all particles. In the following, a derivation of the theoretical solution is provided.
We begin by assuming a steady state, well-developed unidirectional ($\boldsymbol{u}=(u,0,0)$) axial flow driven by a body force $F$. 
Eq.\,\eqref{momentum_equation} simplifies to
\begin{equation}
   \frac{\partial \sigma_{xy}}{\partial y}=-F.   \label{momentumbalance_in_channel}
\end{equation}
The constitutive equation for the only non-zero component of the stress tensor is $\sigma_{xy}=\eta\frac{du}{dy}$
subject to no-slip and vanishing stress boundary conditions $u=0$ at $y=0$ and $\sigma_{xy}=0$ at $y=h$.

Integrating eq.\,\eqref{momentumbalance_in_channel}
with appropriate boundary conditions yields
\begin{equation}
  \sigma_{xy}=F(h-y). \label{stress_in_channel}
\end{equation}

Using eqs.\,\eqref{Maron–Pierce} and \eqref{jamming_point}, the constitutive relation can be rewritten to yield an ODE:
\begin{multline}
 \frac{du}{dy}= F\eta_{\mathrm{s}} (h-y)
 \left(1-  
 \phantom{
\frac{\phi}{\phi_{\mathrm{m}} \exp(-[\frac{\sigma^*}{F(h-y)}]^{\beta})+\phi_0(1-\exp(-[\frac{\sigma^*}{F(h-y)}]^{\beta})}
  }
 \right.  \\ 
 \left. 
 \frac{\phi}{\phi_{\mathrm{m}} \exp(-[\frac{\sigma^*}{F(h-y)}]^{\beta})+\phi_0(1-\exp(-[\frac{\sigma^*}{F(h-y)}]^{\beta})} \right). \label{numericalODE}
\end{multline}

Eq.\,\eqref{numericalODE} can be solved numerically to obtain shear rate, velocity and viscosity profiles. In this work, the solution was obtained via ODE45 MatLab algorithm with a relative tolerance of $10^{-8}$.

\subsubsection{Numerical Results}

\begin{figure}[tb!h]
\centering
\subfloat[][]{\includegraphics[width=0.24\textwidth]{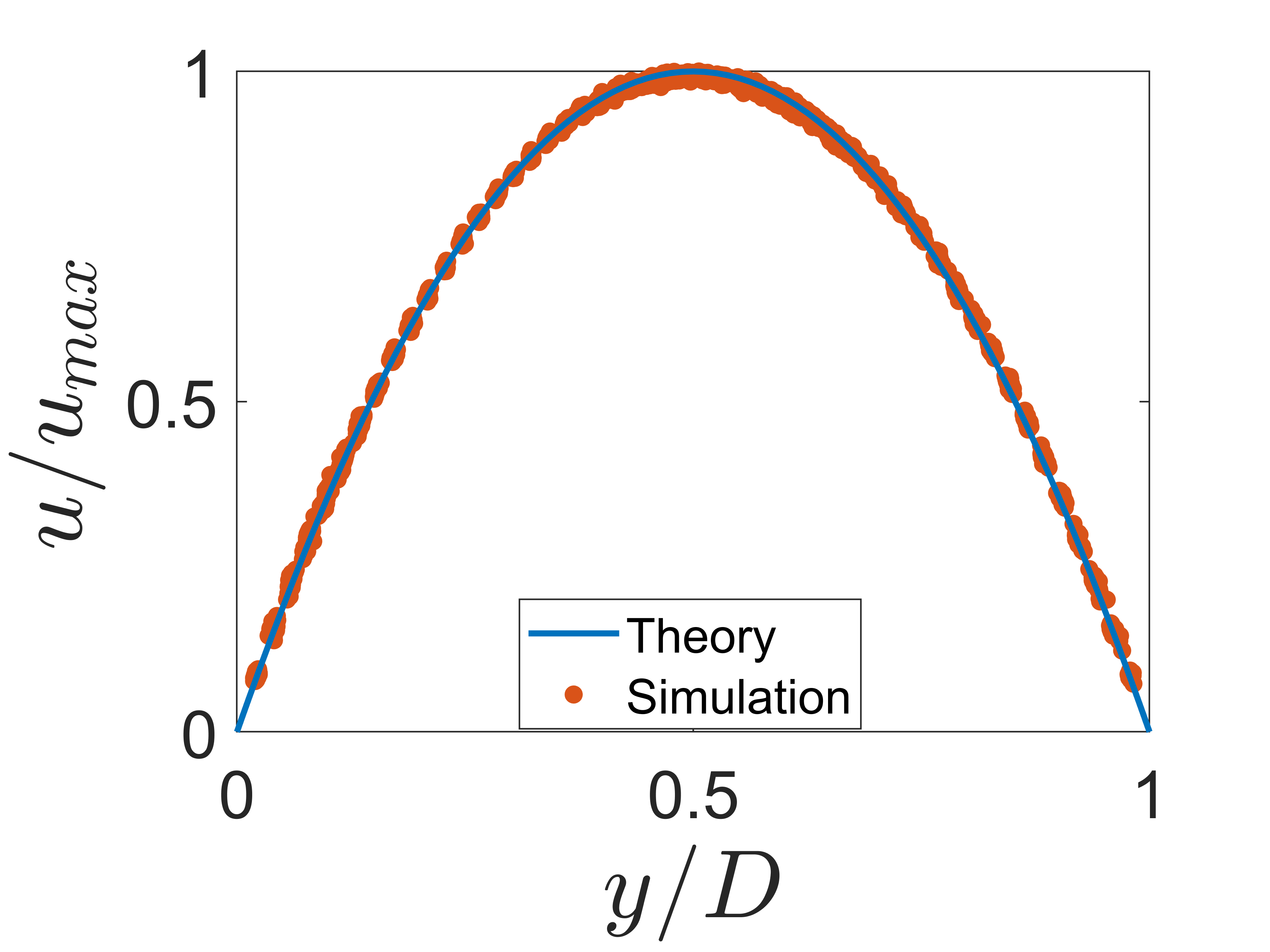}}
\subfloat[][]{\includegraphics[width=0.24\textwidth]{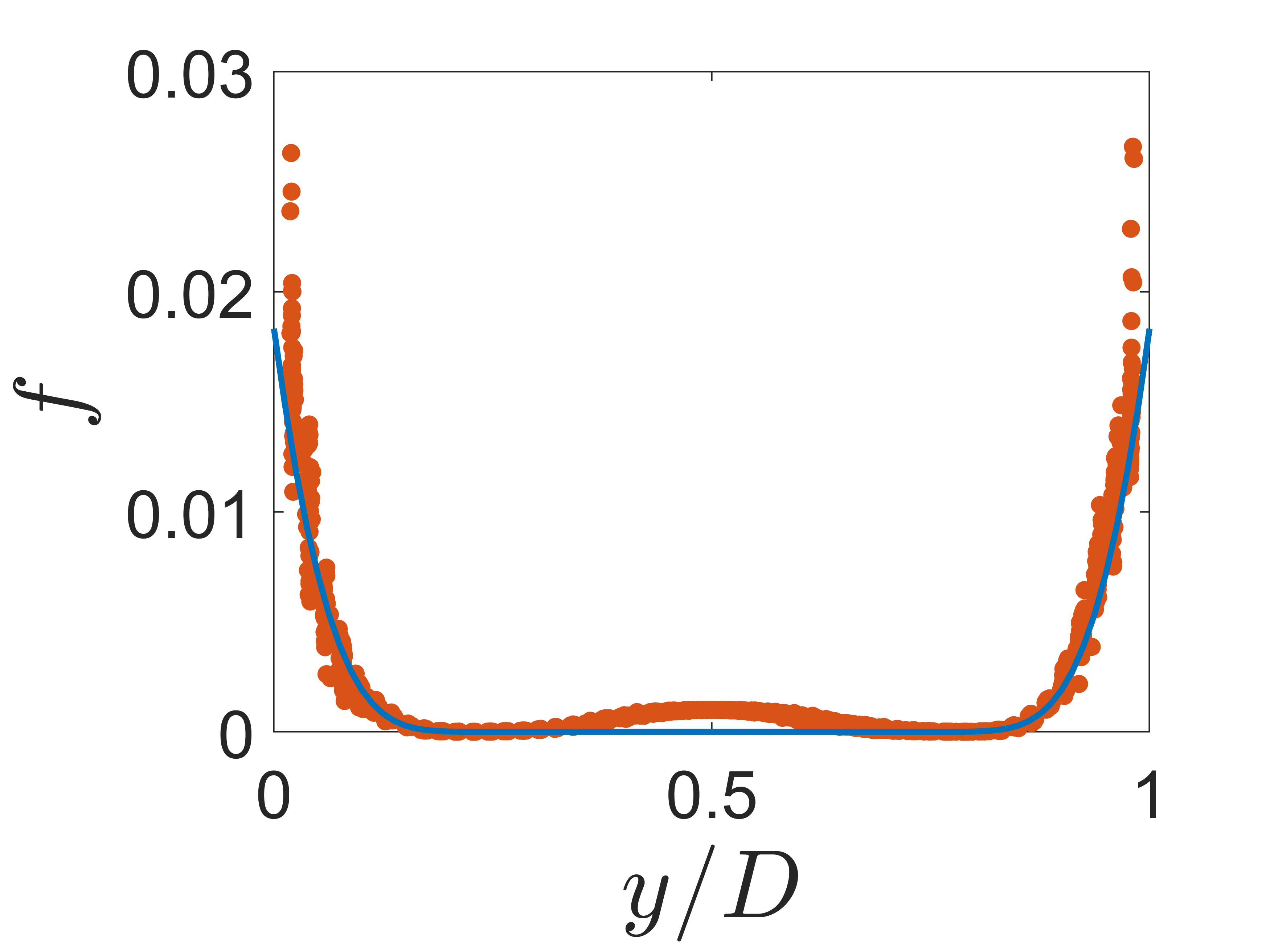}}
 	
\subfloat[][]{\includegraphics[width=0.24\textwidth]{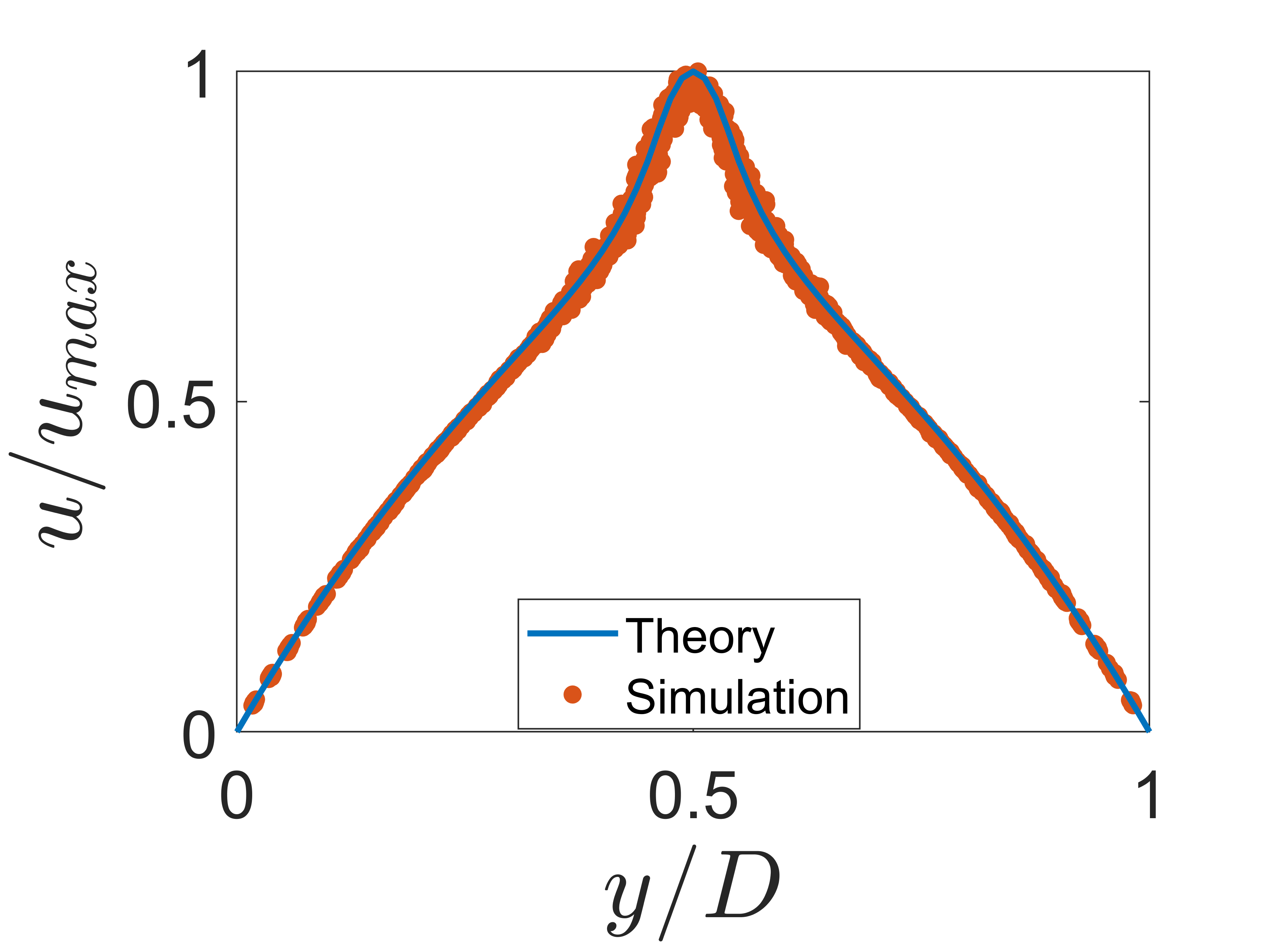}}
\subfloat[][]{\includegraphics[width=0.24\textwidth]{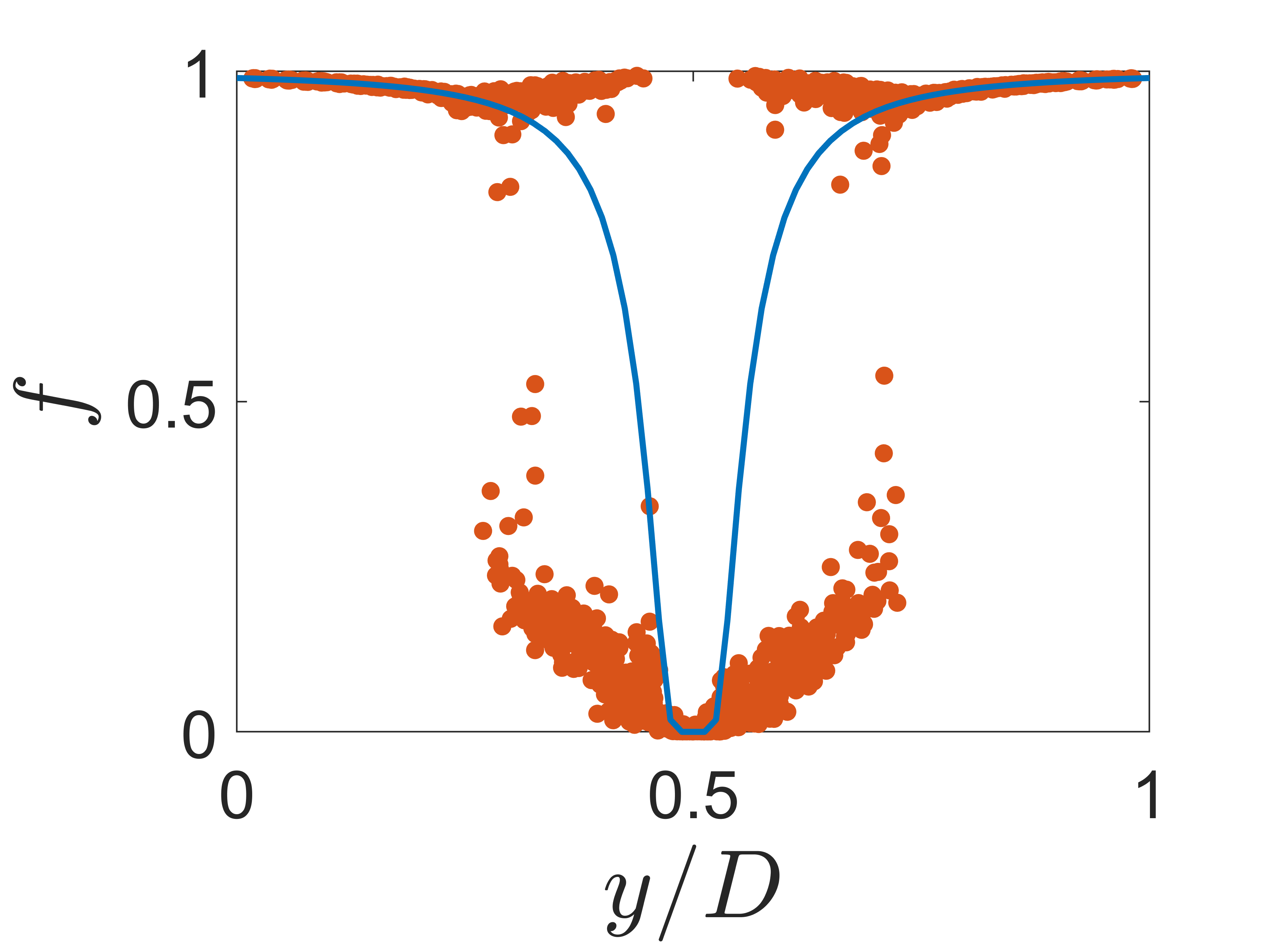}}
 	
\subfloat[][]{\includegraphics[width=0.24\textwidth]{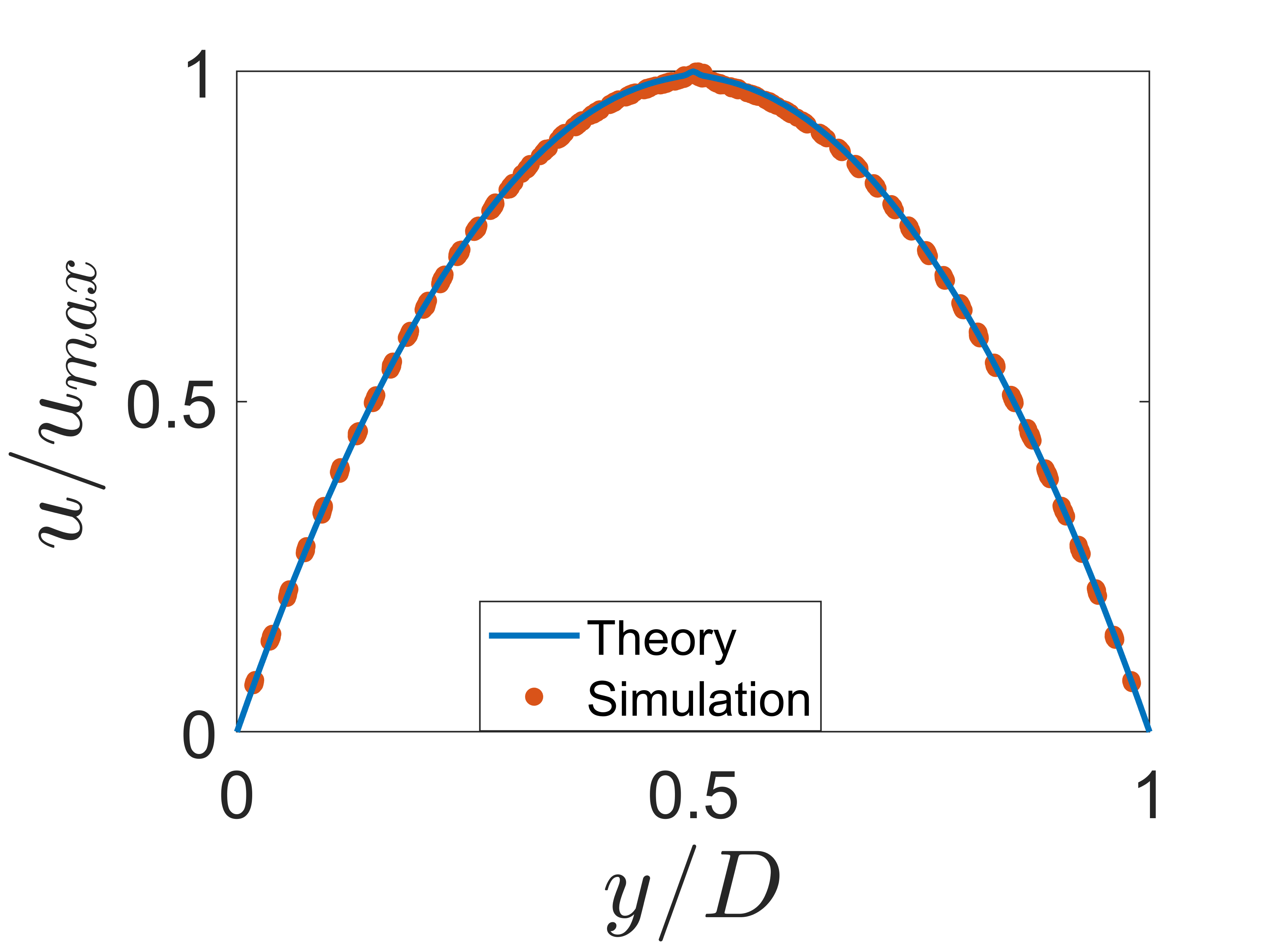}}
\subfloat[][]{\includegraphics[width=0.24\textwidth]{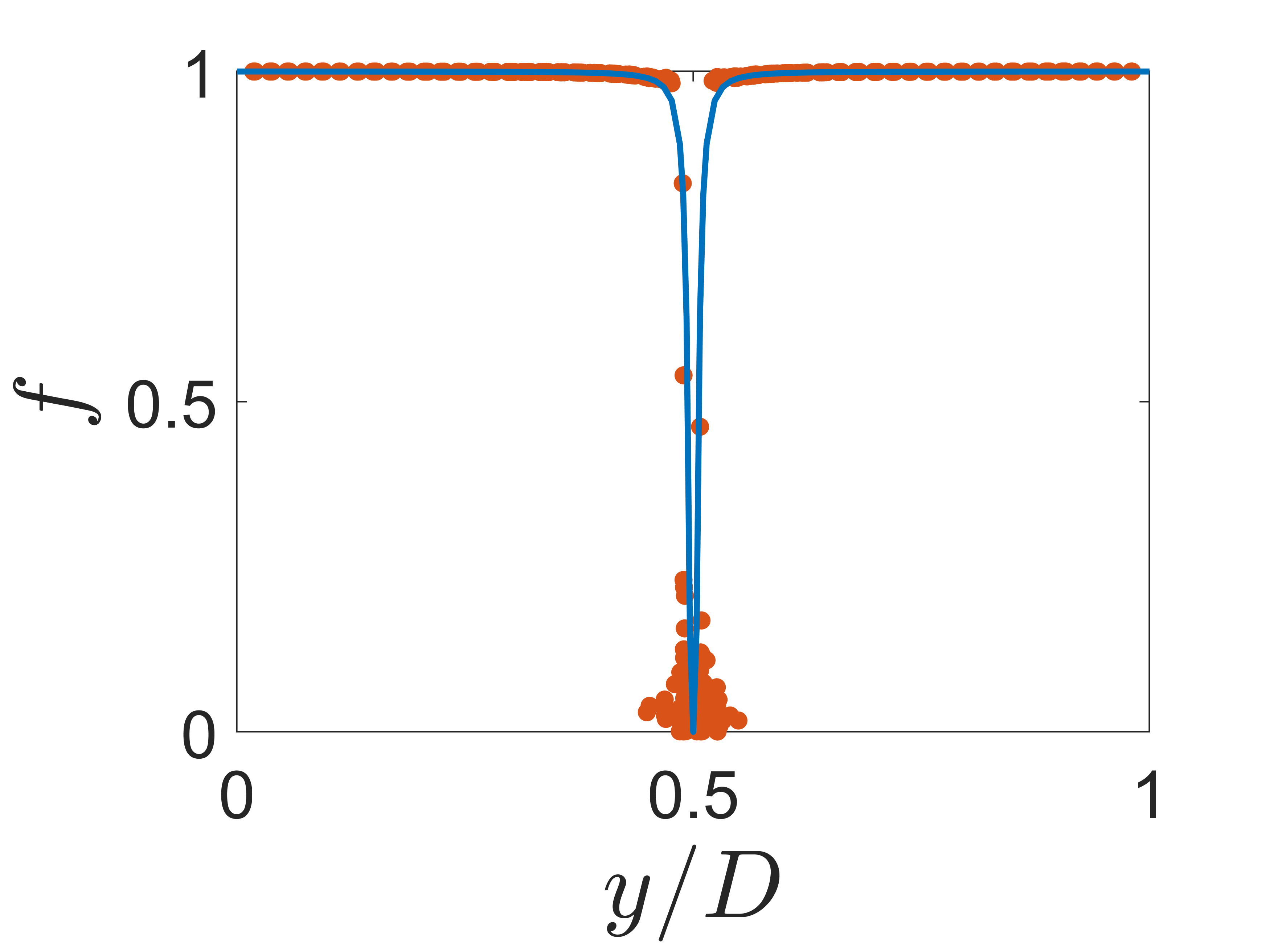}}
\caption{\label{Velocity_MS_profiles}%
Velocity profiles (left) and corresponding microstructural profiles (right) for dimensionless driving forces $Fh/\sigma^*$ of 
0.5 (a)--(b), 10 (c)--(d), and 100 (e)--(f) for $\phi=0.54$.}
 \end{figure}

With the chosen parameters, we focus primarily on transitional DST regime where both frictional and frictionless states are meaningfully present within the flow and consider only the purely local case.
At the steady state, the velocity shape of DST flow displays a sharper profile with an upward inflection near the surface (Fig.\,\ref{Velocity_MS_profiles}(c)) and are in good agreement with the theoretical solutions of our model. 
The inflection corresponds to the threshold transition region between primarily frictional and frictionless regimes. 


It is worth noting that for these values, the ratio between frictionless and frictional viscosity, $\Bar{\eta}$, is approximately 30 --- far lower than typical magnitudes observed in experimental model systems such as cornstarch, where the viscosity can span multiple orders of magnitude during shear thickening\,\cite{Peter2016, Ozturk2020}. This low value of $\overline{\eta}$ was chosen to facilitate feasible simulation run time by avoiding excessively low or high viscosity at any point during shear thickening.

The resulting velocity profiles are presented 
in Fig.\,\ref{Velocity_MS_profiles}. 
The driving force (and thus the stress profile) was increased to probe the influence of shear thickening on the shape of the velocity profiles. 

For sufficiently low driving forces, the velocity profile remains parabolic (a) due to lack of developed microstructure (b) --- viscosity remains close to the lower branch viscosity $\eta_0$. The microstructural profile is in good agreement with the theoretical profile, with minor deviation near the centre line. 
This is due to the initialisation of the structure from a finite value --- particles near centre line experience near-zero shear rate, and thus the structure can not evolve to the appropriate steady state. This deviation is not significantly reflected in the velocity profile due to the weak scaling of viscosity with the microstructure at such low values, resulting in an excellent agreement in the velocity profile.


At intermediate driving forces, flow in the centre ($y/D=0.5$) remains unconstrained; however, the microstructure builds up from the walls (d) towards the centre leading to a sharpened velocity profile (c) with an upward inflection. 
It is apparent that the stress-splitting is present here, with large deviations in the simulated microstructure from the theoretical values. 
Despite this discrepancy, the simulated velocity profile has very good agreement with the theory. 
The banding behaviour seems to preserve average velocity, stress, and microstructural profiles within the channel.


Further increase in the driving force leads to almost completely shear frictional flow (f) --- the velocity becomes broadly parabolic (e) again with the velocity corresponding to the upper branch viscosity $\eta_1$, however, small inflection persists at the tip, where the transition takes place over a very thin region. Any banding behaviour is constrained the small transitional slice, leading to a deteriorated microstructural solution near the centre line.

\begin{figure}[tbh!]
     \centering
     \includegraphics[width=0.45\textwidth]{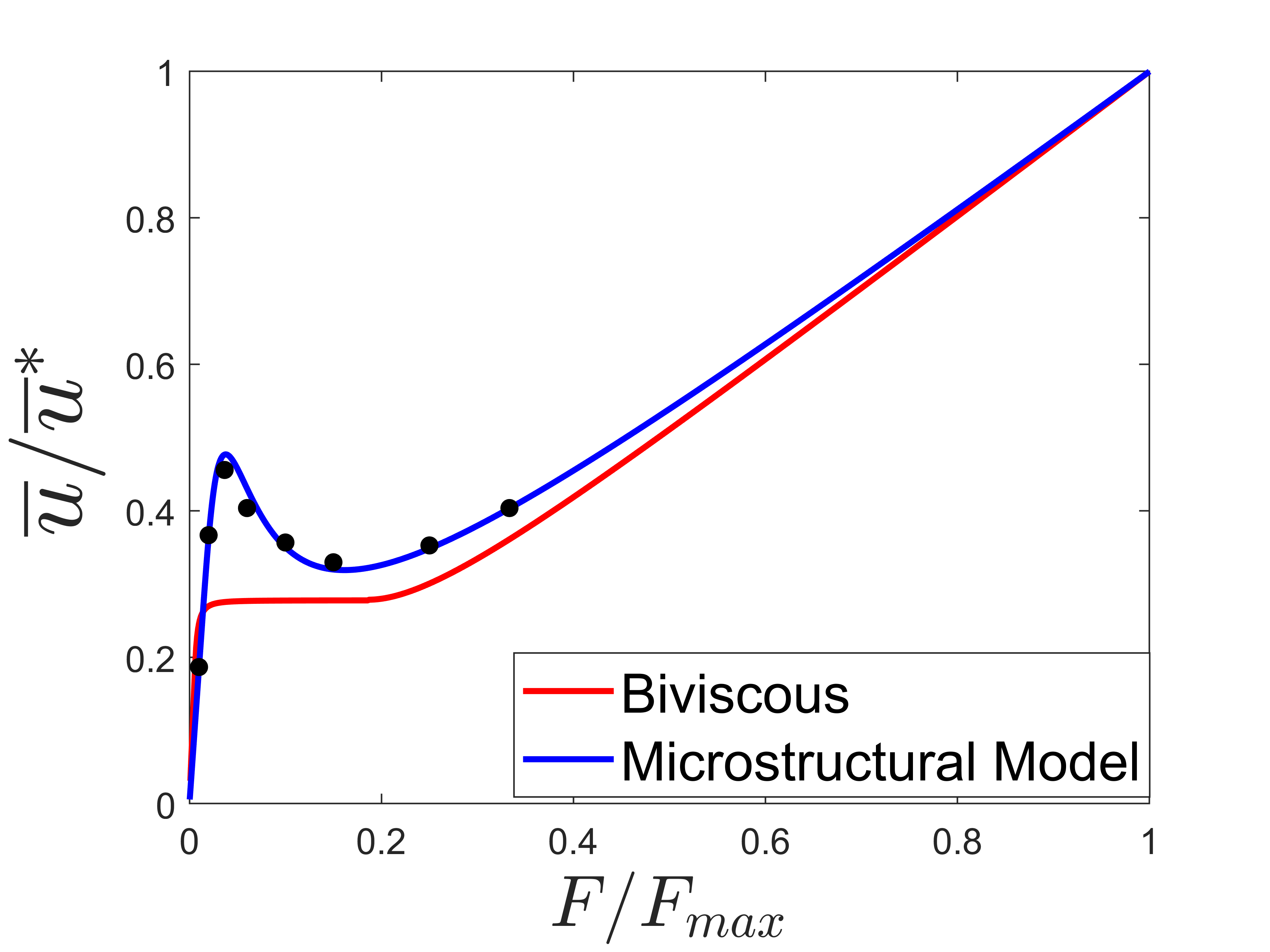}
     \caption{Comparison between the present model (blue) and biviscous model\,\cite{Wagner2017} (red). Simulations (black) carried out for $\phi=0.54$, $\phi_0=0.64$, and $\phi_{\mathrm{m}}=0.56$, $\overline{u}$ is the average velocity in the channel, and $\overline{u}^*$ is an arbitrary average velocity chosen at $F/F_{\mathrm{max}}=1$.}
     \label{comparison_to_biviscous_model}
 \end{figure}
 
The model was compared to the simplest DST model, i.e., 
inverse biviscous\,\cite{Wagner2017}. The biviscous model does not predict S-curves; 
$d\gamma / d\sigma >0$ for all cases. The two models were compared by plotting the average velocity for a range of driving forces 
(Fig.\,\ref{comparison_to_biviscous_model}) for an equivalent viscosity ratio. 
The biviscous model predicts two branches corresponding to lower and upper viscosities, joined together by a plateau where increasing the driving force does not affect the average flow. The model presented in this work exhibits similar low and high viscosity branches; however, the conjoining region exhibits a flow reduction pattern unique to the S-shape rheology. 
It is noteworthy that the presence of a negative flow curve gradient is not always sufficient on its own --- a certain minimum viscosity ratio is required to achieve flow reduction behaviour.

\section{Conclusions}

In this work, we have presented the first implementation of a microstructural model of 
discontinuous shear thickening in SPH simulations. 
A simple non-local scalar microstructural evolution equation based on the work of 
\citet{Kamrin2019} 
was proposed and implemented in simple shear and channel geometries. Measured transients in shear rate were found to be in qualitative agreement with experimentally measured shear rates under constant imposed stress.  

The simple shear results were used to construct flow curves and compare the macroscopic results with the microscopic WC model which we employ.
Simulation results were in good agreement along lower and upper viscosity branches and predicted the characteristic `S-shape' transition between them. 
In the absence of non-locality stress-splitting instability has been observed as a result of the action of the local term in the microstructure evolution equation upon small numerical perturbations in the shear rate field. 
The resulting steady state was found to be characterised by correct domain-averaged shear stress, despite all SPH particles occupying stable upper and lower branches, lack of spatial correlation, and a low shear rate relative to the WC model. 
Both low shear rate and the discrepancy between theoretical and measured split ratios were contextualised in terms of the particle rearrangment mechanism resulting from the stress-control scheme.  
Both moderate and strong non-locality yielded good agreement in average shear rate and shear stress, with banding spatial correlation being observed in the case of moderate non-locality

Channel flow simulations showed excellent agreement in velocity profiles with the theoretical solution of the WC model across a range of driving forces. Once channel flow is forced into the intermediate DST regime, the velocity profile forms an upward inflection as a result of the S-shaped rheology. 
Stress-splitting was observed in the channel flow, with no discernable impact on the velocity profiles. Comparison in channel flow was made with the inverse biviscous mode\,\cite{Wagner2017}. 
The present S-shape microstructure model produced a flow reduction behaviour, in contrast to the flow plateau displayed by the biviscous model.

Further study is required to assess stability of the present model and implementation. 
Exhaustive assessment of the stress-splitting phenomena, band spatial correlation, and the associated characteristic length scale is required. 
Reproduction of known instability phenomena\,\cite{Richards,Fielding} along with introduction of depth in the vorticity direction could provide further insight into the chaotic rheology of the DST materials. 

\begin{acknowledgments}
We thank Zhongqiang Xiong for discussions.
PA and BS acknowledges funding from the Engineering and Physical Sciences Research council [EP/S034587/1].
This research is partially supported by the Basque Government through
the BERC 2022–2025 program and by the Spanish State Research Agency
through BCAM Severo Ochoa excellence accreditation CEX2021-0011
42-S/MICIN/AEI/10.13039/501100011033 and through the project
PID2020-117080RB-C55 (``Microscopic foundations of softmatter experiments: computational nano-hydrodynamics'' and acronym ``Compu-Nano-Hydro''). 
R.S. acknowledges funding from the National Natural Science Foundation of China (12174390, 12150610463) and Wenzhou Institute, University of Chinese Academy of Sciences (WIUCASQD2020002).
\end{acknowledgments}
The data that support the findings of this study are available from the corresponding author upon reasonable request. 
\bibliography{bib1}
\end{document}